\documentclass[twocolumn, tighten, twocolappendix]{aastex631}

\newcommand{\new}[1]{{\color{black} #1}}
\newcommand{\nuwe}[1]{{\color{black} #1}}

\usepackage{rotating}
\usepackage{comment}
\usepackage[caption=false]{subfig}
\usepackage{enumerate}
\usepackage{float}

\newcommand{\Mjup}{{\rm M}_{\rm Jup}}

\newcommand{\Mth}{{\rm M}_{\rm th}}

\newcommand{\Msun}{{\rm M}_\odot}

\newcommand{\Lsun}{{\rm L}_\odot}

\newcommand{\Mp}{M_{\rm p}}

\newcommand{\rpdisk}{R_{\rm p}}
\newcommand{\rpsky}{r_{\rm p}}
\newcommand{\PApdisk}{\phi_{\rm p}}
\newcommand{\PApsky}{{\rm PA}_{\rm p}}

\newcommand{\Mstar}{M_{\star}}

\newcommand{\Lstar}{L_{\star}}
\newcommand{\Tdust}{T_{\rm d}}

\newcommand{\xo}{\Delta x}
\newcommand{\yo}{\Delta y}
\newcommand{\cspeed}{c_{\rm s}}
\newcommand{\Sigmag}{\Sigma_{\rm gas}}

\newcommand{\CO}{^{12}{\rm CO}}
\newcommand{\tCO}{^{13}{\rm CO}}

\newcommand{\agrain}{a_{\rm grain}}
\newcommand{\St}{\rm St}

\newcommand{\sigmaSB}{\sigma_{\rm SB}}

\defcitealias{pinte19-hd97048}{Pin19}
\defcitealias{pinte20-dsharpkinks}{Pin20}

\defcitealias{izquierdo22-hd163296}{Izq22}
\defcitealias{verrios22-imlup}{Ver22}

\defcitealias{pinte18-hd163296}{Pin18}
\defcitealias{rosotti21-citau}{Ros21}
\defcitealias{bae22-AS209-CPD}{Bae22}

\graphicspath{{./}{figures/LO-RES}} 

\received{2022 September 20}
\accepted{2022 November 1}
\submitjournal{\apjl}

\shorttitle{Midplane Evidence for the Velocity Kink Planet Candidates?}
\shortauthors{Speedie \& Dong}

\begin{document}

\title{\textbf{Testing velocity kinks as a planet-detection method:\\ Do velocity kinks in surface gas emission trace planetary spiral wakes in the midplane continuum?}}

\correspondingauthor{JS, RD}
\email{jspeedie@uvic.ca, rbdong@uvic.ca}

\author[0000-0003-3430-3889]{Jessica Speedie}
\affiliation{Department of Physics \& Astronomy, University of Victoria, Victoria, BC, V8P 1A1, Canada}

\author[0000-0001-9290-7846]{Ruobing Dong}
\affiliation{Department of Physics \& Astronomy, University of Victoria, Victoria, BC, V8P 1A1, Canada}
\affiliation{Kavli Institute for Astronomy and Astrophysics, Peking University, Beijing 100871, China}

\begin{abstract}
Spiral density waves generated by 
an embedded planet are understood to cause ``kinks'' in observed velocity channel maps of CO surface  
emission, by perturbing the gas motion within the spiral arms. 
If velocity kinks are a reliable probe of embedded planets, we should expect to see the planet-driven spiral arms in other observational tracers.
We test this prediction by searching the dust continuum for the midplane counterparts of the spirals responsible for all of the velocity kink planet candidates reported to date, whose orbits lie inside the \new{dust} continuum disk. 
\new{We find no clear detection of any spiral structure in current continuum observations for 6 of the 10 velocity kink planet candidates in our sample} 
(DoAr 25, GW Lup, Sz 129, HD 163296 \#2, P94, and HD 143006), despite the high planet masses inferred from the kink amplitude. 
\new{The remaining 4 cases include 3 clear detections of two-armed dust spirals (Elias 27, IM Lup and WaOph 6) wherein neither spiral arm aligns with a wake originating from reported planet location, suggesting that under the planetary-origin hypothesis, an accurate method for inferring the location of the planet in the midplane may need to encompass vertical effects.} 
The 10th case, HD 97048, is inconclusive with current knowledge of the disk geometry.

\end{abstract}


\section{Introduction} \label{sec:introduction}

The velocity ``kink'' kinematic signature has gained standing as a promising method for discovering embedded planets \citep{diskdynamicscollab20, pinte22-ppvii}.
To date, 12 planet candidates have been reported based on velocity kink detections; in three cases, the planets have been ingested into the NASA or European exoplanet databases as \textit{confirmed} planets\footnote{\href{https://exoplanets.nasa.gov/exoplanet-catalog/7503/hd-97048-b/}{HD 97048 b}; \href{http://exoplanet.eu/catalog/hd_163296_b/}{HD 163296 b}; \href{http://exoplanet.eu/catalog/hd_163296_c/}{HD 163296 c} (links embedded). } \citep{pinte19-hd97048, izquierdo22-hd163296}, and in one case, the candidate's circumplanetary disk has been observed co-located with the kink \citep[AS 209,][]{bae22-AS209-CPD}. 
Many more such detections of embedded planets are expected with the upcoming \texttt{exoALMA} Large Program\footnote{2021.1.01123.L; \href{https://www.exoalma.com/}{https://www.exoalma.com/}}.

Some loose ends exist, however, that motivate independent verification of the planetary origin of velocity kinks.
Eleven of the 12 velocity kink detections were made in $\CO$ emission, which is expected to originate above the midplane \citep[e.g.,][]{law21-maps4-emissionsurfaces}, outside of where any analytic theory of velocity kinks has been achieved \citep{bollati21-theory-of-kinks-2d}.
Ten of the 12 detections were made by visual inspection, without an assessment of the statistical significance of the kink signal \citep{pinte18-hd163296, pinte19-hd97048, pinte20-dsharpkinks}, and in some cases the detections do not appear in independent datasets \citep{teague21-maps18-hd163296-mwc480}. 
While the planet hypothesis is on the one hand supported by the inferred planet locations coinciding with dust gaps, the mass needed to generate kinks with the observed amplitudes is \new{higher than} the planet mass derived from the properties of the dust gaps by a factor of $4-100$ \cite[e.g.,][]{lodato19-newbornplanets-dustgaps, zhang18-dsharp7-planetdiskinteractions}. 

The velocity kink signal is understood to be generated by the embedded planet's spiral wakes. Along the spiral arms, the gas 
motion is perturbed 
relative to Keplerian rotation, which appears as an excess and absence of emission (i.e., a ``kink'') in the channel maps \citep[e.g., ][]{bollati21-theory-of-kinks-2d}. 
This understanding enables us to make a robust, testable prediction: \textit{All instances of planet-driven velocity kinks should be concurrent with planet-driven spiral arms.} 
\new{Recently, \citet{calcino22-hd163296} demonstrated that the velocity kinks observed in $\CO$ emission in HD 163296 map directly onto the theoretical curve for the spiral wake driven by planet c, projected up onto the emission surface.} 
In this letter, we search \new{the disk midplane for the spiral arms driven by 10 velocity kink planet candidates, using (sub-)mm continuum observations, under the usual assumption that the $\sim$(sub)-mm-sized dust traced by such observations has settled to the disk midplane}. Our goal is to provide an independent verification for the existence of the predicted planets, and thereby test the validity of velocity kinks as signposts of planets in disks.

\section{Data \& Methods} \label{sec:methods}

\subsection{Sample: Disks with Velocity Kinks} \label{subsec:sample}

We compile all the disks with velocity kinks reported in the literature to date: 
\begin{itemize}
    \item 1 kink in HD 163296 (``HD 163296 \#1'') from \citet{pinte18-hd163296} (hereafter \citetalias{pinte18-hd163296}) 
    \item 1 kink in HD 97048 from \citet{pinte19-hd97048} (hereafter \citetalias{pinte19-hd97048})
    \item 9 kinks (8 new) in 8 DSHARP disks: Elias 27, HD 143006, HD 163296 (a second kink in this disk: ``HD 163296 \#2''), IM Lup, DoAr 25, GW Lup, Sz 129 and WaOph 6, from \citet{pinte20-dsharpkinks} (herafter \citetalias{pinte20-dsharpkinks})
    \item 2 kinks in HD 163296 (a third unique kink, ``P94'', and an independent re-detection of  HD 163296 \#1, dubbed ``P261'') from \citet{izquierdo22-hd163296} (hereafter \citetalias{izquierdo22-hd163296})
    \item and 1 kink in AS 209 co-located with a CPD candidate from \citet{bae22-AS209-CPD} (hereafter \citetalias{bae22-AS209-CPD}).
\end{itemize}
In total, 12 velocity kinks have been reported in 10 disks. With the exception of P261 and P94 in HD 163296 \citepalias[which were identified by \texttt{discminer};][]{izquierdo22-hd163296}, all the detections have been made by visual inspection of the channel maps, and the statistical significance of the detections has not been quantified. We list the disks and relevant properties of the kink detections in Table \ref{tab:kinks}.

We then exclude from our sample the detections with inferred planet locations exterior to the outer edge of continuum emission. This eliminates HD 163296 \#1 a.k.a P261 \citepalias{pinte18-hd163296, izquierdo22-hd163296}
and AS 209 \citepalias{bae22-AS209-CPD}, leaving us with a total of 10 velocity kinks in 9 disks (where the repeat disk is HD 163296 containing \citetalias{pinte20-dsharpkinks}'s HD 163296 \#2 kink and \citetalias{izquierdo22-hd163296}'s P94 kink).

As a side note, velocity deviations attributed to a planet have also been reported in HD 100546 \citep{casassus19-hd100546, perez20-hd100546} and TW Hya \citep{teague22-twhya-kinematics}. We do not consider these detections in this work primarily\footnote{
Additionally, the HD 100546 planet lies inside a continuum ring \citep[$0.01 \pm 0.04 \arcsec$, $0.21 \pm 0.04 \arcsec$ on the sky;][]{casassus19-hd100546}, which is contrary to the classical paradigm that embedded planets carve gaps \citep[though would support the scenario of][]{nayakshin20-twhya}.
Since the initial discovery, the velocity deviations in HD 100546 have been explained as being due to an inner binary companion \citep{norfolk22-hd100546}, and due to disk eruptions driven by an embedded outflow \citep{casassus22-hd100546-eruption}.
The inferred planet location in TW Hya ($\rpsky=1.53 \arcsec$ or 82 au, $\PApsky=60^{\circ}$) lies outside of the outer edge of continuum emission that is detected when observed at high angular resolution \citep[$\sim 30$ mas,][]{huang18-twhya}, and the existing observations with sufficient sensitivity to detect continuum emission extending beyond 82 au are too low angular resolution for our purposes \citep[$0.37 \arcsec$,][]{ilee22-twhya-deep}.} 
because we are concerned with detections based on kinks in velocity channel maps, whereas these were made based on Doppler flips in velocity residual maps.

\textit{Planet location.} The works reporting the velocity kinks in our sample provide the midplane location of the predicted planets. As we are testing this prediction, we adopt the given locations. For 9 of our 10 candidate planets \citepalias[the exception being P94 in HD 163296;][]{izquierdo22-hd163296}, the procedure that was used to determine the planet location is the following \citepalias{pinte19-hd97048, pinte20-dsharpkinks}: 
\begin{enumerate}[i)]
    \item Identify the CO channel in which the velocity kink is most prominently detected, by visual inspection;
    \item Identify the center of the kink in that channel, by visual inspection;
    \item Measure the altitude of the CO emission surface at the center of the kink, using the method of \citet{pinte2018-altitude-of-CO};
    \item Deproject that location onto the disk midplane.
\end{enumerate} 
In these 9 cases, the planet position is given in sky-coordinates ($\rpsky$, $\PApsky$) without an estimate of spatial uncertainty. 

The location of the P94 planet in HD 163296 is retrieved by \texttt{discminer},\footnote{We note that \texttt{discminer} assesses the significance of a deviation from Keplerian velocity, not whether the deviation matches the expected morphology of a planet-driven velocity deviation (as this has not yet been described analytically in 3 dimensions).} in disk-frame coordinates ($\rpdisk$, $\PApdisk$), with an uncertainty in the radial and azimuthal directions of $\pm 6$ au and $\pm 3$ degrees, respectively \citepalias{izquierdo22-hd163296}. However, due to \texttt{discminer}'s velocity centroid folding procedure,\footnote{A way of removing contributions to the velocity field that are symmetric about the disk minor axis, stemming from gas gaps and bulk disk rotation.} the retrieved polar angle is degenerate 
about the disk minor axis and additional information or reasoning is needed to subsequently determine if the detection is on the redshifted or blueshifted side of the disk. The \texttt{discminer} velocity residuals of P94 were found to have a Doppler flip morphology, and by reasoning that the sub-Keplerian branch should 
be interior to the planet's orbit and the super-Keplerian branch should be exterior \citep{bollati21-theory-of-kinks-2d}, \citetalias{izquierdo22-hd163296} report the planet on the redshifted side. 
We note that the ``mirror kink'' (i.e., the detection with opposite sign but equal significance, on the blueshifted side) is co-located with  \citetalias{pinte20-dsharpkinks}'s HD 163296 \#2 kink \citepalias[Footnote 11,][]{izquierdo22-hd163296}.

We list the reported locations of the candidate planets in our sample in Table \ref{tab:planets}. 
In all cases, the inferred location pinpoints the planet within a dust gap.

\subsection{Dataset: Continuum Observations} \label{subsec:observations}

The continuum data we present in this Letter are from the same ALMA program as the CO data in which the velocity kinks were detected. 
    For the 8 \citetalias{pinte20-dsharpkinks} disks, we retrieve the publicly available, self-calibrated science-ready continuum images and \new{fiducial (continuum-subtracted)} $\CO \, J=2-1$ image cubes  
    from the DSHARP data repository.\footnote{\href{https://almascience.eso.org/almadata/lp/DSHARP/}{https://almascience.eso.org/almadata/lp/DSHARP/}} For HD 97048, we obtain the self-calibrated continuum images and $\tCO \, J=3-2$ cube from Figshare.\footnote{\href{https://figshare.com/articles/dataset/HD_97048_ALMA_B7_continuum_13CO/8266988}{DOI: https://doi.org/10.6084/m9.figshare.8266988.v1}}
Our analysis is only focused on the continuum images, and we do no new analysis on the CO cubes.
Table \ref{tab:observations} summarizes the observations and some basic properties of the data. For observational setup and data reduction details, we refer the reader to
\citetalias{pinte19-hd97048} for HD 97048, and to \citet{andrews18-dsharp1} for the 8 DSHARP disks in our sample. We measure the rms noise in the continuum images inside an annulus centered on the disk whose outer radius is the maximum allowed by the field of view and inner radius sufficiently larger than the source, following \citet{andrews18-dsharp1}. For reference, we also measure the rms noise in the CO cubes, in the same annular area, throughout the first and last 5 channels.

\subsection{Methods: Searching for Dust Spirals} \label{subsec:ingredients}

\textit{Approach.} Since spirals are perturbations in surface brightness above/below the background disk, our approach is to subtract an axisymmetric model for the background in order to extract the spiral signal. We create this axisymmetric background model in the image plane, by azimuthally averaging the continuum image. 

\nuwe{We choose to do our analysis in the image plane, rather than the visibility domain, for two main reasons. Firstly, it yields similar results to $uv$-plane fitting in terms of both the morphology and sensitivity of the residuals, and is easily reproducible. We discuss this in more detail in Appendix \S\ref{app:subsec:frank-andrews}. Secondly,} the disk conditions (equation of state, optical depth, planet mass) and observing setups (angular resolution, sensitivity) under which \nuwe{the image plane} method successfully retrieves the spiral signal has been quantified on synthetic continuum observations of planet-driven dust spirals in hydrodynamic simulations \citep{speedie22-alma-dustspirals}. \nuwe{We therefore can form apples-to-apples expectations for the observability of the dust spirals, which we describe in detail in \S\ref{subsec:expectations}.}

\textit{Disk geometry.} 
Knowledge of the inclination and position angle of the continuum disk 
is needed to create the axisymmetric background model, and Table \ref{tab:disks} provides the geometrical parameters we use for each disk. 
For 4 of the 8 DSHARP disks (HD 143006, HD 163296, GW Lup and Sz 129), we use the values found in \citet{andrews21-frank-dsharp} (their Table 2) by the \texttt{frank} \citep{jennings20-frank} residual appearance method. \new{We also tried the \citet{huang18-dsharp2} geometries for these 4 disks and found that it had no effect on our results.} 
For the other 4 not in \citet{andrews21-frank-dsharp} (Elias 27, IM Lup, DoAr 25 and WaOph 6), we use the values found in \citet{huang18-dsharp2} (their Table 2) by fitting ellipses to individual annular dust substructures. 
For HD 97048, no continuum-derived geometry has been published to our knowledge. We thus adopt two possible geometries, found by different methods: 
(i) fitting a Keplerian disk model to the velocity field from CO line data cubes \citep[Table C.1,][]{bohn22-misalignments} using \texttt{eddy} \citep{teague19-eddy}; 
(ii) fitting ellipses to rings and gaps in near-IR scattered light \citep[Table 1,][]{ginski16-hd97048-sphere}.

\textit{Disk rotation direction.} We assume the predicted planet orbits in the same direction that the disk rotates. For HD 163296, DoAr 25 and HD 97048, it has been determined which side of the disk major axis is the near/far side with existing scattered light observations (see notes for Col. 10 in Table \ref{tab:disks} for references), and we use that information in conjunction with knowing which side about the minor axis is the blue/redshifted side to deduce the direction that the disk rotates. For HD 143006 and GW Lup (low inclination disks), the near/far side determination not definitive \citep[e.g.,][]{benisty18-hd143006-SL}, but \citet{perezL18-hd143006} suggest HD 143006's west side is the near side, and \citet{garufi22-sphere} posit GW Lup's northwest side is mostly likely the near side. This would mean HD 143006 rotates clockwise and GW Lup rotates anti-clockwise. In the former case this happens to be the opposite direction to the ``low-level'' tentative large-scale Archimedean spiral found by \citet{andrews21-frank-dsharp} \new{(see Appendix \S\ref{app:subsec:frank-andrews} for more discussion)}. For Sz 129, a relatively unstudied disk, no scattered light observations exist in the literature to our knowledge. Our results (Figs. \ref{fig:sec3:non-detections-1} \& \ref{fig:sec3:non-detections-2}) will show that the rotation direction of these 3 disks --HD 143006, GW Lup and Sz 129-- is rendered irrelevant by the lack of spiral features in their continuum residual maps, but we still wish to show the tightness of the spiral winding. For that purpose, we assign clockwise for HD 143006 and anti-clockwise for GW Lup \citep[motivated by the suggestions of][]{perezL18-hd143006, garufi22-sphere} and anti-clockwise Sz 129 (arbitrarily). 
For Elias 27, IM Lup and WaOph 6, we adopt the rotation direction found in \citet{huang18-dsharp3-spirals} (their \S3.2). 

\begin{rotatetable*}
\begin{deluxetable*}{cccccccccccc}
\tablenum{1}
\tablecaption{Sample and summary of possibly planet-induced velocity kink detections to date. \label{tab:kinks}}
\tablewidth{0pt}
\tabletypesize{\scriptsize}
\tablehead{
\colhead{Disk} & \colhead{Ref.} & \colhead{Method} & \colhead{Line} & \colhead{$v_{\rm res}$} & \colhead{$v_{\rm kink \, channel}$} & \colhead{$\Delta v$} & \colhead{$\sigma_{\rm kink}$} & \colhead{S/N$_{\rm CO}$}  & \colhead{Gap} & \colhead{$M_{\rm p}$} & \colhead{Notes}\\
\colhead{ } & \colhead{ } & \colhead{ } & \colhead{ } & \colhead{(m/s)} & \colhead{(km/s)} & \colhead{ } & \colhead{ } & \colhead{ }  & \colhead{ } & \colhead{($\Mjup$)} & \colhead{}
}
\decimalcolnumbers
\startdata 
\multicolumn{12}{c}{Kinks inside the continuum emission disk} \\ \hline
Elias 27        & \citetalias{pinte20-dsharpkinks} & VI & $\CO (2-1)$  & 350   & 1.70  & ``?''                          & Firm      & 12      & D69  & $1-3$ & \textit{a} \\
HD 143006       & \citetalias{pinte20-dsharpkinks} & VI & $\CO (2-1)$  & 320   & 8.84 & $\approx 0.2 \, v_{\rm Kep}$    & Firm      & 10      & D22  & $1-3$ & \textit{g} \\
HD 163296 (P94) & \citetalias{izquierdo22-hd163296}  & \texttt{discminer} & $\CO (2-1)$ & 320  & 6.28 & 0.41 km/s       & (19.4, 7.5)  & ...  & D86  & $1-3$ & \textit{b}     \\
HD 163296 (\#2) & \citetalias{pinte20-dsharpkinks} & VI & $\CO (2-1)$  & 320   & 3.40  & $\approx 0.15 \, v_{\rm Kep}$  & Firm      & 36      & D86  & $1-3$ & \textit{b, f}     \\
IM Lup          & \citetalias{pinte20-dsharpkinks} & VI & $\CO (2-1)$  & 350   & 3.05 & $<0.24 \, v_{\rm Kep}$          & Firm      & 14      & D117 & $1-3$ & \textit{c}     \\
DoAr 25         & \citetalias{pinte20-dsharpkinks} & VI & $\CO (2-1)$  & 350   & 5.05 & ``?''                           & Tent.      & 7      & D98  & $1-3$ & \textit{a} \\
GW Lup          & \citetalias{pinte20-dsharpkinks} & VI & $\CO (2-1)$  & 350   & 2.70  & $<0.3 \, v_{\rm Kep}$          & Tent.      & 12     & D74  & $1-3$ & \\
Sz 129          & \citetalias{pinte20-dsharpkinks} & VI & $\CO (2-1)$  & 350   & 4.80  & $<0.2 \, v_{\rm Kep}$          & Tent.      & 11     & D64  & $1-3$ & \\
WaOph 6         & \citetalias{pinte20-dsharpkinks} & VI & $\CO (2-1)$  & 350   & 2.10  & ``?''                          & Tent.      & 13     & D79  & $1-3$ & \textit{a} \\
HD 97048        & \citetalias{pinte19-hd97048}     & VI & $^{13}{\rm CO} (3-2)$ & 120   & 5.76 & ...                    & ...      & ...      & 130 au  & $2-3$ & \\\hline
\multicolumn{12}{c}{Kinks outside the continuum emission disk} \\ \hline
HD 163296 (\#1, P261) & \citetalias{pinte18-hd163296}   & VI \& \texttt{discminer} & $\CO (2-1)$  & 110   & 1.00    & 0.40 km/s                 & (5.2, 4.6)$^{d}$ & ...     & N/A  &  2 & \textit{d, e}     \\
AS 209          & \citetalias{bae22-AS209-CPD}     & VI & $\CO (2-1)$  & 200  & 4.80  & ...                  & ...      & ...        & N/A   &  $1.3 \cdot (\alpha / 10^{-3})^{1/2}$ &   \\
\enddata
\tablecomments{The last two rows are not in our sample, as the inferred planet location lies outside the continuum emission disk, but we include them for completeness. Column descriptions:\\
(1) Name of disk. Name of kink or planet candidate in brackets, if applicable.\\
(2) Paper first reporting the velocity kink. All values in the corresponding row are from this reference unless otherwise noted.\\
(3) Method by which the velocity kink was detected. ``VI'' means visual inspection of the channel maps, and \texttt{discminer} is the quantitative tool of \citet{izquierdo21-discminer}.\\
(4) CO isotopologue and $J$ transition in which the kink is reported.\\
(5) Velocity (spectral) resolution of the CO cube in which the kink is reported.\\
(6) Velocity (relative to Earth) of the channel in which the kink is most prominently detected, which is used to pinpoint the planet location (see Sec. \ref{subsec:sample}). The two exceptions to this are: \textit{(i)} \citetalias{izquierdo22-hd163296} (row 3): the value in this entry is one of two channels in which the authors note the kink can be seen visually (caption of their Fig. 1), and \texttt{discminer} is used to pinpoint the planet location; \textit{(ii)} \citetalias{bae22-AS209-CPD} (row 12): the value in this entry is the channel in which the CPD candidate is most clearly detected in $\tCO$, and is the central of three channels in which the $\CO$ velocity kink is reported.\\ 
(7) Amplitude of the velocity deviation. ``?'' is verbatim from the reporting paper, and `...' means not provided by the authors. For an independent velocity deviation prediction from 3D simulations for some of the DSHARP disks, see \citet{rabago-zhu-21}. \\
(8) Either a qualitative classification as a ``firm'' or ``tentative'' kink detection by \citetalias{pinte20-dsharpkinks}, or, the statistical significance ($\sigma_r, \sigma_\phi$) of the deviation from Keplerian velocity from \texttt{discminer} \citepalias{izquierdo22-hd163296}.\\
(9) Signal to noise of CO emission at the location of the kink.\\
(10) Dust gap associated with inferred planet location. Gap name designation from \citet{huang18-dsharp2} for the DSHARP disks, and approximate gap radius in au for HD 97048 \citepalias{pinte19-hd97048}.\\ 
(11) Mass estimate of the candidate planet, inferred from the velocity kink amplitude. For the \citetalias{pinte18-hd163296}, \citetalias{pinte19-hd97048} and \citetalias{pinte20-dsharpkinks} disks, this is from forward modeling with SPH simulations; for P94 it is from forward modeling with hydrodynamic simulations \citepalias{izquierdo22-hd163296}; and for AS 209 \citepalias{bae22-AS209-CPD} it is from the \citet{kanagawa16-planetmass-gapwidth} empirical relation between the gas gap width and planet mass. For mass estimates derived from the dust gap properties, see \citet{zhang18-dsharp7-planetdiskinteractions} and \citet{lodato19-newbornplanets-dustgaps}.\\
(12) Notes:\\
\textit{(a)} Channel maps suffer from cloud contamination \citep[Column 7, Table 5,][]{andrews18-dsharp1}. Visually, the affected velocities are: $2.75-4.85$ km/s (7 channels) for Elias 27; $1.55-5.05$ km/s (11 channels) for DoAr 25, and $2.45-4.20$ km/s (6 channels) for WaOph 6.\\ 
\textit{(g)} Simulation work supporting the existence of this planet: \citet{ballabio21-hd143006}.\\
\textit{(b)} The HD 163296 \#2 and P94 kinks have the same polar angle, if mirrored about the disk minor axis \citepalias[Footnote 11,][]{izquierdo22-hd163296}. \\
\textit{(c)} Simulation work supporting the existence of a planet in this disk: \citet{verrios22-imlup}.\\
\textit{(d)} Value of $\sigma_{\rm kink}$ in this row is from \citetalias{izquierdo22-hd163296} with \texttt{discminer}.\\
\textit{(e)} Simulation work supporting the existence of this planet: \citet{calcino22-hd163296}. \\
\textit{(f)} Not verified in $\CO$ channel maps from the MAPS program \citep{teague21-maps18-hd163296-mwc480}.
}
\end{deluxetable*}
\end{rotatetable*}


\subsection{Expectations: Dust Spirals Driven by the Velocity Kink Planets} \label{subsec:expectations}

\textit{Midplane spiral morphology.} We expect the embedded planets predicted by the velocity kink detections to
drive spiral wakes in the gas at the midplane whose intrinsic morphology (amplitude, width and phase) is determined by the planet mass and location, as well as disk temperature.

As we are searching for these spirals in the (sub-)mm continuum and not the gas, the first question is whether we expect a difference between the dust spiral morphology and the morphology of the spiral in the gas. 
This depends on how quickly the dust responds to the change in aerodynamic drag forces exerted by the gas when the grains encounter the gas spiral perturbation, which in turn depends on the dust grain size and the local gas surface density. \citet{sturm20-dustspirals} and \citet{speedie22-alma-dustspirals} showed that for dust with Stokes number ($\St \propto \agrain \, \Sigmag^{-1}$, where $\agrain$ is the dust grain size and $\Sigmag$ is the local gas surface density) lower 
than the critical Stokes number, $\St \lesssim \St_{\rm crit} \approx 0.05-0.1$,\footnote{The critical Stokes number is the Stokes number for which the time it takes a dust grain to cross the spiral wake is equal to the grain's stopping time, and so this range is introduced by the azimuthal width of the wake, which changes with planet mass and distance from the planet \citep[\S3.2,][]{speedie22-alma-dustspirals}.} the dust responds quickly enough such that the resulting dust spiral is morphologically identical to the driving gas spiral at the midplane. We expect our continuum observations to be most sensitive to thermal emission from dust grains of size $\agrain \approx \lambda_{\rm obs} /2 \pi$ \citep{kataoka15-dustscattering, pavlyuchenkov19-spectralindex}, which translates to $\agrain \approx 0.14$ mm for HD 97048 (Band 7) and $\agrain \approx 0.20$ mm for the 8 DSHARP disks (Band 6). 
For gas surface densities higher than just $\sim0.2 \, {\rm g \, cm^{-2}}$, these grain sizes correspond to Stokes numbers lower 
than $\St_{\rm crit}$ \citep[see Fig. 2 of][]{speedie22-alma-dustspirals}.
\nuwe{Figure 7 of \citet{dullemond18-dsharp6} shows inferred gas surface densities for a subset of the DSHARP disks (including 3 in our sample) to range  $0.1 \lesssim \Sigmag \lesssim 50\, {\rm g\, cm^{-2}}$.}
\nuwe{To put it another way, assuming a gas surface density profile $\Sigmag \sim 1/r$, then in order for 
$\Sigmag$ 
to be lower than $0.2\, {\rm g\, cm^{-2}}$ at 50 au (typical location of inner arms in our sample), the total disk mass contained within $100$ au would need to be lower than $6.7 \times 10^{-4} \, \Msun$.} We therefore expect no difference between the intrinsic morphology of the predicted \nuwe{midplane} gas spiral and that of the dust spiral we aim to observe, and can use the literature knowledge of gas spirals to understand the morphology of the expected dust spirals. 

The trajectory 
of a planet-driven spiral (i.e., the azimuthal location of the spine, or peak amplitude, as a function of radius) is the result of constructive interference among 
various spiral wave modes, each excited by a different Fourier component of the planet's gravitational potential \citep{bae+zhu18-spirals1, bae+zhu18-spirals2}. To predict where we expect to see positive residuals (emission above the axisymmetric background) in the continuum residual map for each planet in our sample, we use the analytic phase equation of \citet{bae+zhu18-spirals1}:\footnote{This equation assumes a circular orbit for the planet; see \citet{zhu22-eccentric-spirals} and \citet{fairbairn22-eccentricplanet-spirals} for semi-analytic linear theory of spiral density waves excited by planets on eccentric orbits.}
\begin{eqnarray}
\label{eqn:bae+zhu-spiral}
    \phi_{m, n}(R) = &-& \PApdisk - {\rm sgn}(R - \rpdisk) { \pi \over 4m} + 2\pi  {n \over m} \\
    &-& \int_{R_m^\pm}^{R} {\Omega(R') \over c_s(R')} \left| \left(1- {R'^{3/2} \over {\rpdisk^{3/2}}} \right)^2 - {1 \over m^2} \right|^{1/2} {\rm d}R' \nonumber \, ,
\end{eqnarray}
where ($\rpdisk$, $\PApdisk$) are the midplane coordinates of the planet in the disk frame, $\Omega(R)$ is the angular velocity of the disk, $\cspeed(R)$ is the sound speed of the gas, and $m$ is the azimuthal wavenumber of the wave mode excited by the $m$th Fourier component of the planet's potential, which itself has a number of components indexed by $n$. The $n=0$ components form the primary spiral arms, which are easier to recognize than e.g. secondary arms ($n=1$ for $R<\rpdisk$, or $n=m-1$ for $R>\rpdisk$) because: (i) they are launched relatively near to the planet, at Lindblad resonances $R_m^\pm = (1 \pm 1/m)^{2/3}R_{\rm p}$ \citep{goldreich-tremaine79}, 
and therefore the inner and outer primary arms always ``point'' to the planet, whereas the location of the additional arms (both the starting point and the azimuthal separation from the primary) varies with planet mass \citep{fung-dong15-inferring-planet-mass-spirals}; 
and (ii) close to the planet, they have the highest amplitude \citep{bae+zhu18-spirals1}. As such, we set $n=0$. 

The third term in Eqn. \ref{eqn:bae+zhu-spiral} is the only radially-varying term, and describes how tightly wound the spiral wave modes are as they propagate away from the planet. In addition to $m$, this term depends on the gas pressure scale height, $H(R) = \cspeed / \Omega$. We calculate $\Omega(R)$ as the Keplerian angular velocity $\Omega(R) = (G \, \Mstar / R^3)^{1/2}$, where $R$ is the disk-frame radial coordinate and $\Mstar$ is the stellar mass (Col. 3, Table \ref{tab:disks}). We calculate the gas sound speed as
\begin{equation}
    \cspeed(R) = \left( {k_{\rm B} \, \Tdust(R) \over \mu \, m_{\rm prot} } \right)^{1/2} \, ,
    \label{eqn:cs}
\end{equation}
where $k_{\rm B}$ is the Boltzmann constant, $\mu=2.37$ is the mean molecular weight of the gas in atomic units, and $m_{\rm prot}$ is the proton mass. We thus need an analytic estimate for the disk temperature at the midplane $\Tdust(R)$, for which we use the simple irradiated flaring disk recipe of \citet{dullemond18-dsharp6}:
\begin{equation}
    \Tdust(R) = \Bigg(  { {1 \over 2} \varphi \Lstar \over  4 \pi r^2 \sigmaSB } \Bigg)^{1/4} \, .
    \label{eqn:Tmid-dullemond18}
\end{equation}
Here, $\Lstar$ is the luminosity of the central star (Col. 4, Table \ref{tab:disks}), $\sigmaSB$ is the Stefan-Boltzmann constant and $\varphi$ is the flaring angle \cite[e.g.,][]{chiang-goldreich97, dullemond01-irradiated}. A smaller flaring angle corresponds to a colder temperature profile and a more tightly wound spiral. We assign $\varphi=0.02$ to be consistent with \citet{dullemond18-dsharp6} and \citet{huang18-dsharp2}.

Returning to the $m$ dependence of the third term in Eqn. \ref{eqn:bae+zhu-spiral}, we expect the phase of the spiral we see to follow that of the dominant azimuthal mode, $m_{\rm dom}=({1\over2})(H/r)_{\rm p}^{-1}$, in the case of low mass planets \citep[$\Mp \lesssim 0.1 \, \Mth$, where $\Mth = \cspeed / \Omega G = (H/r)_{\rm p}^3\, \Mstar$ is the unit of thermal mass;][]{bae+zhu18-spirals1, bae+zhu18-spirals2}. However, for higher mass planets, the wave modes propagate at faster speeds, and the resulting spiral arms are more open \citep{goodman-rafikov01} and should more closely follow lower ($m < m_{\rm dom}$) modes. 
In Col. 8 of Table \ref{tab:planets}, we convert the predicted masses of the embedded planets ($1-3 \, \Mjup$, Col. 8 of Table \ref{tab:kinks}) into units of $\Mth$ using our estimation of ($H/r$)$_{\rm p}$ (Col. 7 of Table \ref{tab:planets}), and find $\Mp > 1.0 \, \Mth$ in every case. For this reason, we consider azimuthal wave modes down to the lowest possible $m$ ($m=1$ in the outer disk and $m=2$ in the inner disk). We also consider $m\rightarrow \infty$, corresponding to the linear limit of \citet{rafikov02-spirals} and the most tightly wound spirals \cite[used in applications to observations by e.g.,][]{muto12-spirals-hd135344B, casassus21-filament-spirals-hd135344B}.

\textit{Observability.} Using synthetic continuum observations, \citet{speedie22-alma-dustspirals} found that the dust spirals driven by thermal mass planets at 50 au in a slowly cooling and moderately inclined ($i\lesssim 50^{\circ}$) disk $140$ pc away are detectable in continuum observations with sensitivity between $10-25 \, \mu{\rm Jy \, beam ^{-1}}$ and angular resolution $\sim 30-65$ mas. \citet{dong-fung17-bright-spirals-SL} show that the amplitude of the spirals increases with planet mass for sub-thermal mass planets, and flattens out for super-thermal mass planets (their Fig. 1). In our sample, the inferred planet location is at a few tens to $\sim$100 au, the mean beam size is $42 \pm 12$ $\times$ $54 \pm 17$ mas, the mean distance to the source is $144 \pm 26$ pc, the estimated continuum rms noises are all $\leq 22.6\, \mu{\rm Jy \, beam ^{-1}}$ (except HD 97048), and only 2 disks are inclined by greater than $50^{\circ}$. 
\new{We thus expect the current continuum observations to be sensitive to dust spirals driven by planets of thermal mass and above. \nuwe{Using the estimated $(H/r)_{\rm p}$ (Col. 7, Table \ref{tab:planets}) and known $\Mstar$ (Col. 3, Table \ref{tab:disks}) for each candidate in our sample, this $1.0 \, \Mth$ lower limit translates to Jupiter masses ranging between $0.15-0.96 \, \Mjup$ (with $0.15\, \Mjup$ corresponding to HD 143006, and $0.96 \, \Mjup$ to HD 97048).} }

Note that gravitational instability may also produce spiral arms in continuum emission detectable in ALMA residual maps \citep{hall19-GI-spirals-ALMA}, and they may interfere with planet-induced spirals \citep{rowther22-GI-planet-spirals}. We do not account for this complication in this work.

\section{Results \& Discussion} \label{sec:results-discussion}

\begin{figure*}
\begin{center}
\includegraphics[width=18cm]{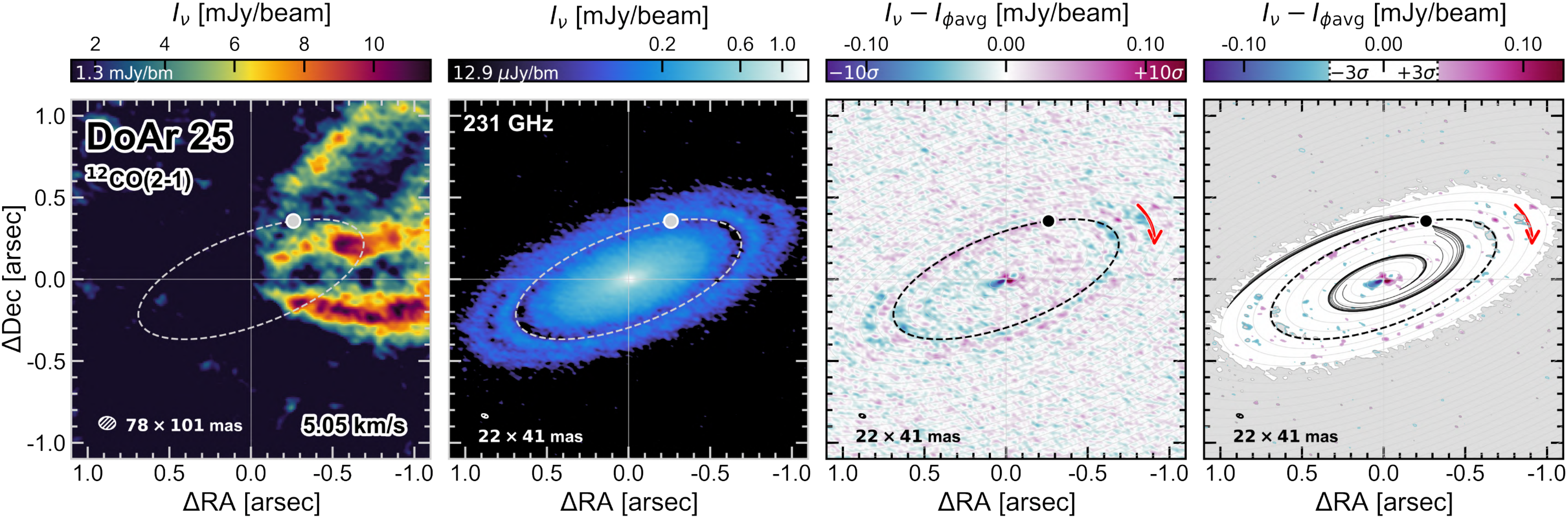}
\includegraphics[width=18cm]{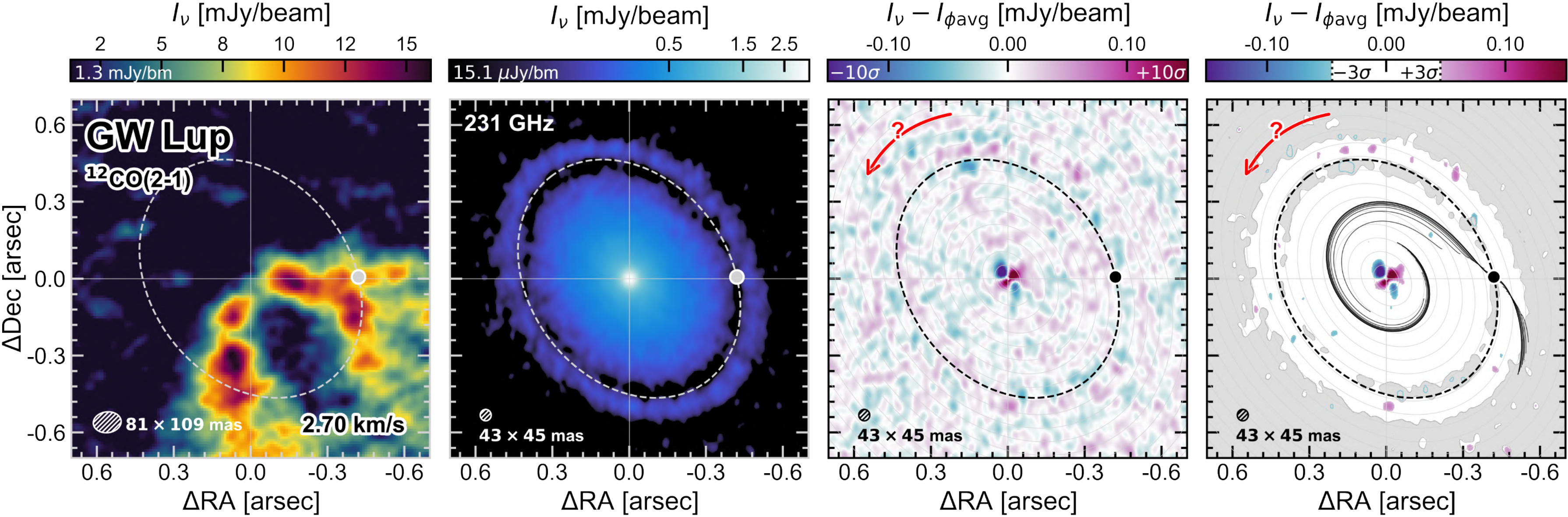}
\includegraphics[width=18cm]{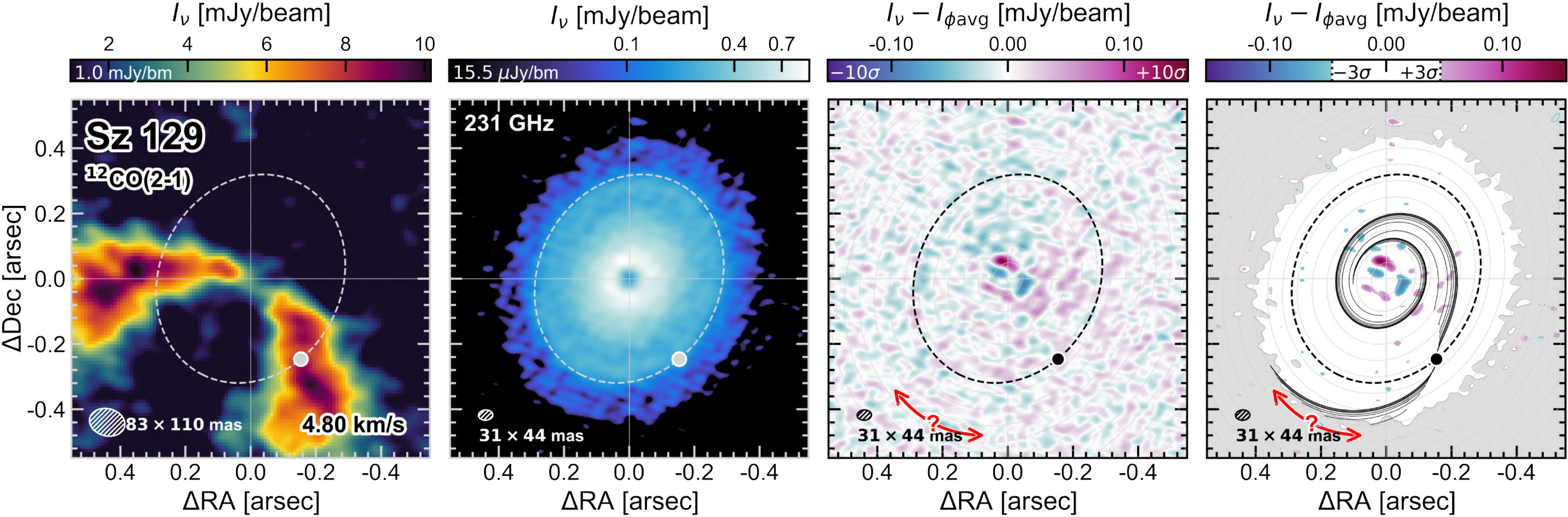}
\end{center}
\caption{\textbf{No clear detections of the predicted dust spirals:} DoAr 25, GW Lup and Sz 129. In all columns, the grey or black point marks the reported planet location in the midplane, and the dashed grey or black line shows its circular orbit.\\ 
\textit{1st column:} The CO channel map in which the velocity kink is most prominently detected. The estimated rms noise in the cube is written in the colorbar, and the colormap starts at that value. \\
\textit{2nd column:} Continuum image showing where the planet lies relative to substructures in the dust distribution. The colormap starts at three times the rms noise (again written in the colorbar) and has a $1\over4$-power law stretch.\\
\textit{3rd column:} Continuum residuals after subtracting the azimuthal average. Red arrows indicate the direction of rotation of the disk, and in all cases the arrow is located at the redshifted major axis. The colorbar spans $\pm10\times$ the rms noise.\\
\textit{4th column:} Comparison between detected residual substructures stronger than $3\times$ the continuum rms noise, and the theoretical prediction for the midplane spiral wake driven by the candidate planet \cite[][our Eqn. \ref{eqn:bae+zhu-spiral}]{bae+zhu18-spirals1, bae+zhu18-spirals2}. Light grey indicates where emission in the continuum image falls below this same threshold, helping to distinguish whether an absence of spiral-shaped residuals is due to the non-presence of spiral, or non-presence of emission (e.g. inside dust gaps or beyond the edge of the disk).
Thin grey ellipses are projected concentric circles in radial steps of 1 beam major axis, helping to discern spirals from circular arcs under the angular resolution of the image.
\label{fig:sec3:non-detections-1}}
\end{figure*}

We find non-detections of dust spirals for 6 of the 10 candidate planets in our sample: DoAr 25, GW Lup, Sz 129, HD 163296 \#2, P94, and HD 143006 (Fig. \ref{fig:sec3:non-detections-1} \& \ref{fig:sec3:non-detections-2}, \S\ref{subsec:non-detections}).
In 3 cases (Elias 27, IM Lup and WaOph 6), dust spirals are detected but their locations do not agree with that of the predicted planet (Fig. \ref{fig:sec3:detections}, \S\ref{subsec:detections}). For the 10th candidate planet, HD 97048, the result is inconclusive (Fig. \ref{fig:sec3:inconclusive}, \S\ref{subsec:inconclusive}).

\subsection{Non-detections} \label{subsec:non-detections}

\begin{figure*}
\begin{center}
\includegraphics[width=18cm]{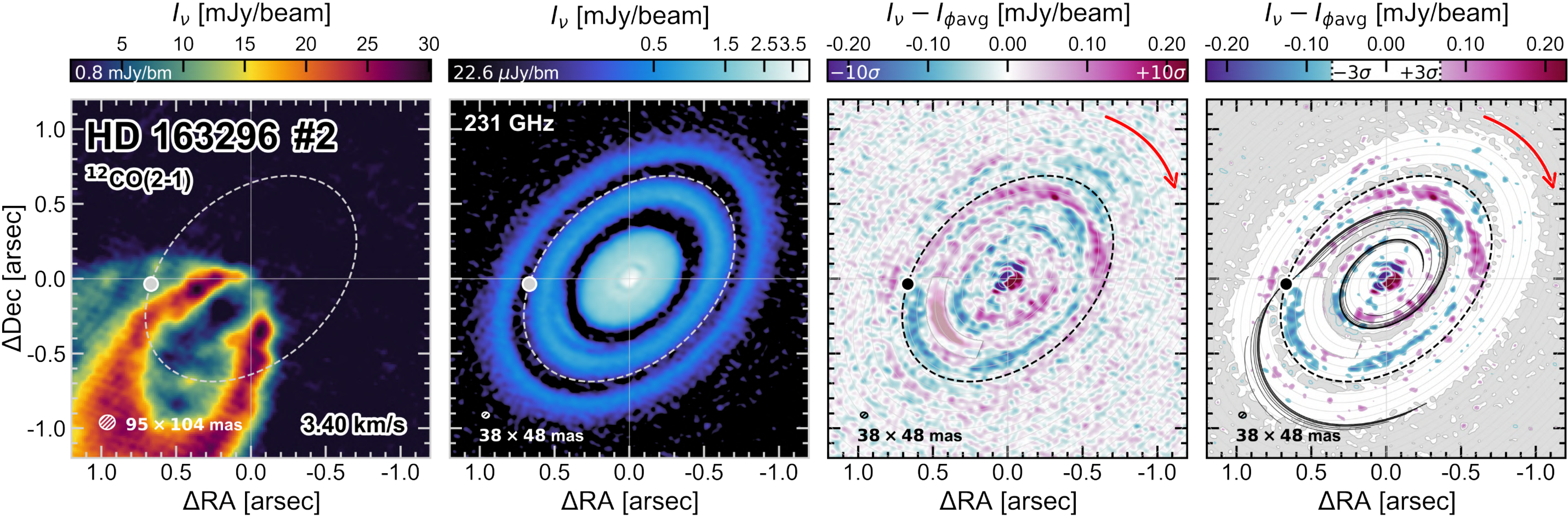}
\includegraphics[width=18cm]{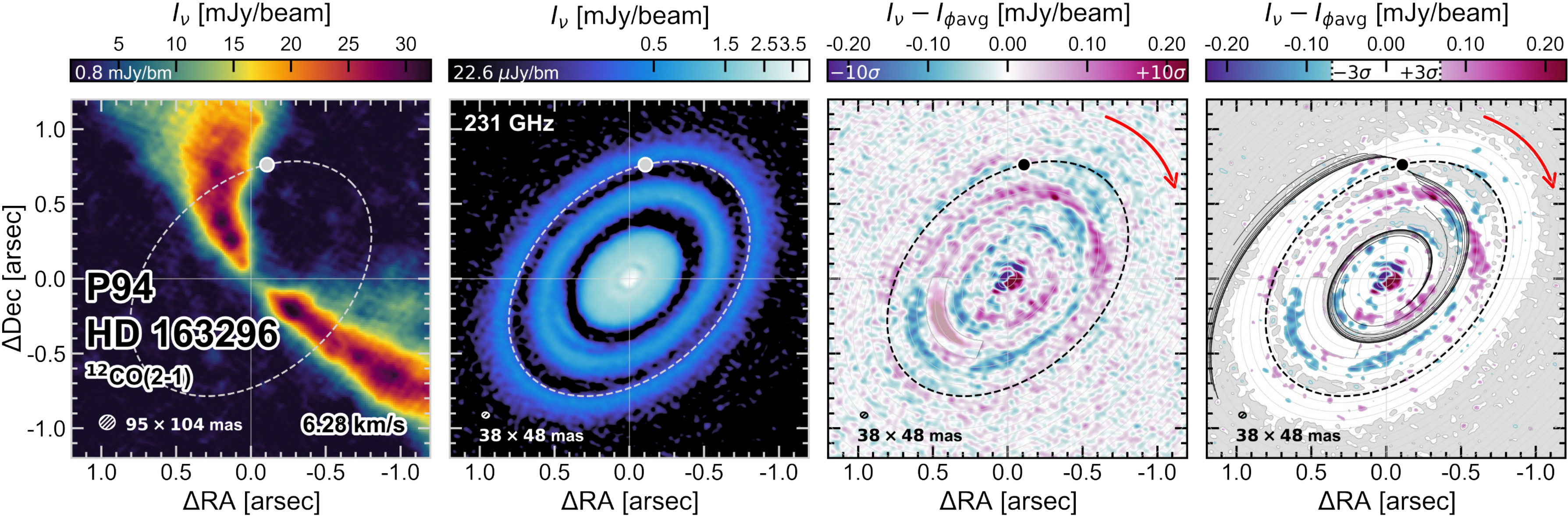}
\includegraphics[width=18cm]{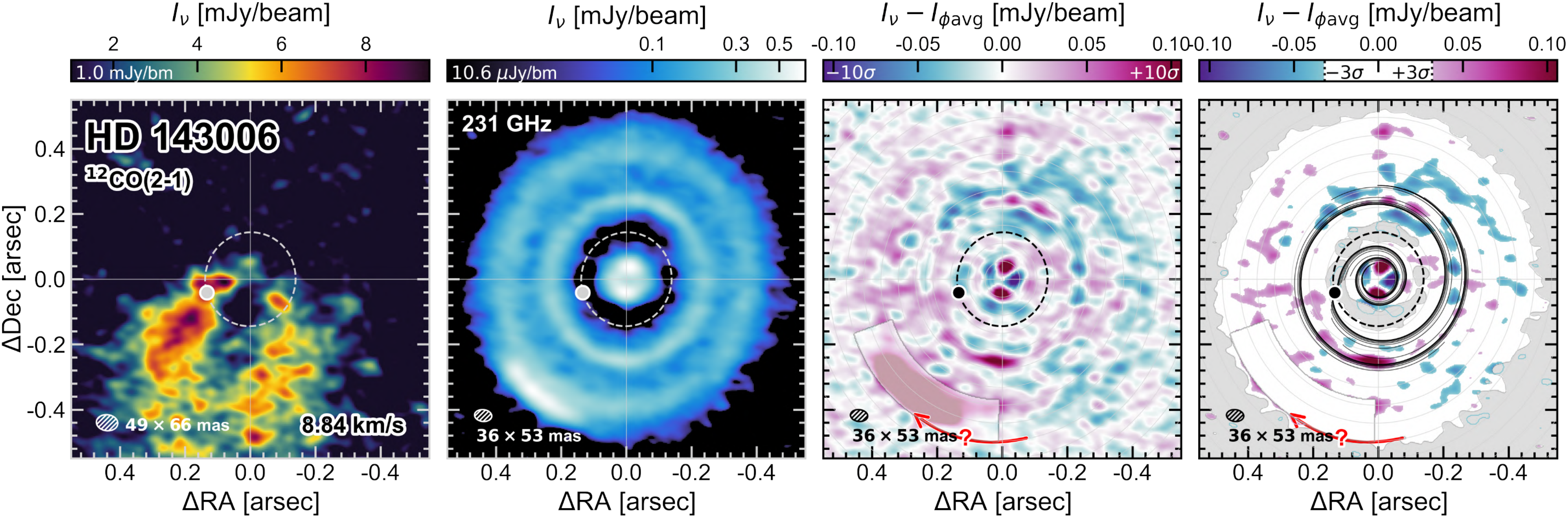}
\end{center}
\caption{\textbf{No clear detections of the predicted dust spirals (continued):} HD 163296 \#2, P94, and HD 143006.\\
We mask the pronounced arc-like azimuthal asymmetries in these two disks to enhance the possibility of spiral detection (see Appendix Fig. \ref{fig:app:masking-asymmetries}).\\
\new{Additional figures showing the continuum residuals after re-imaging the calibrated measurement sets with different Briggs parameters are available for HD 163296, HD 143006, DoAr 25, GW Lup and Sz 129 at \href{https://doi.org/10.6084/m9.figshare.21330426}{DOI: 10.6084/m9.figshare.21330426}.}
\label{fig:sec3:non-detections-2}}
\end{figure*}

\begin{figure*}
\begin{center}
\includegraphics[width=18cm]{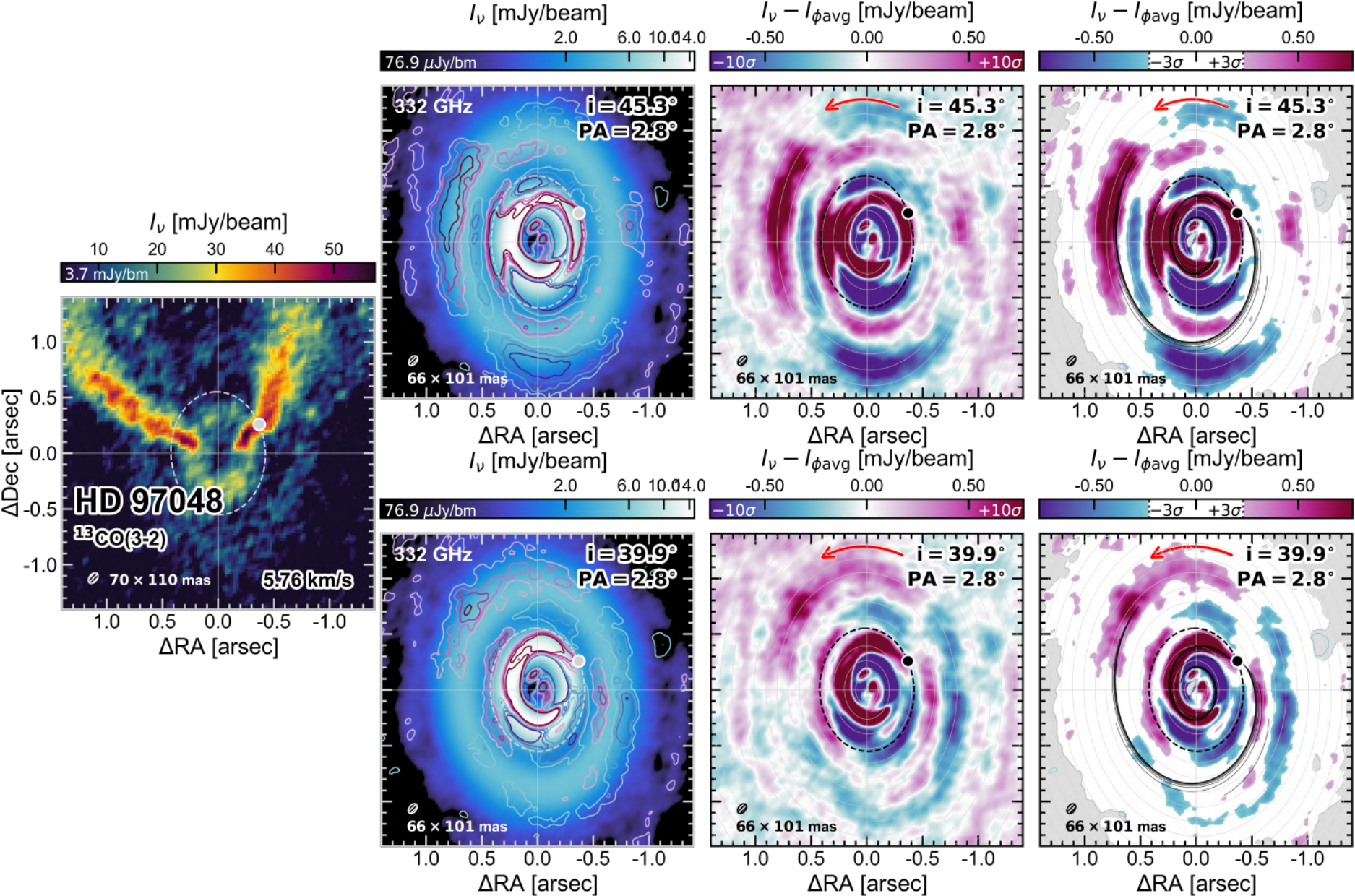}
\end{center}
\caption{\textbf{Inconclusive case:} HD 97048. See \S\ref{subsec:inconclusive} for details. Residual contours of $\pm 3, 5, 7, 10 \times$ the continuum rms noise are overlaid in the 2nd column to help identify where residuals lie in relation to the gap and rings.  
\textit{Top row:} Continuum residuals calculated assuming the disk geometry of \citet{bohn22-misalignments}. 
\textit{Bottom row:} Continuum residuals calculated assuming the disk geometry of \citet{ginski16-hd97048-sphere}. 
Appendix Fig. \ref{fig:app:high-pass-hd97048} provides non-geometry-dependent residual maps for this disk.
\label{fig:sec3:inconclusive}}
\end{figure*}

Of the 6 non-detections of dust spirals, 3 correspond to velocity kinks that were classified as ``tentative'' detections \citepalias[DoAr 25, GW Lup, Sz 129;][]{pinte20-dsharpkinks}. In these disks, we find no significant non-axisymmetric continuum substructure (Fig. \ref{fig:sec3:non-detections-1}). 

Of the latter 3 non-detections (Fig. \ref{fig:sec3:non-detections-2}), 2 correspond to ``firm'' kink detections \citepalias[HD 163296 \#2, HD 143006;][]{pinte20-dsharpkinks}, and 1 to a kink detection with a radial and azimuthal significance of $(\sigma_r$, $\sigma_{\phi}) = ($19.4, $7.5)$ \citepalias[P94; ][]{izquierdo22-hd163296}. We find some small-scale non-axisymmetric continuum substructures in these disks, but none that agree with the predicted spiral wakes. 

\new{The above results persisted in additional imaging efforts we performed with the calibrated measurement sets for DoAr 25, GW Lup, Sz 129, HD 163296 and HD 143006, varying the Briggs parameter to maximize the observing sensitivity (see Appendix \S\ref{app:re-imaging}).}

\textit{If the planets are there, why don't we see the dust spirals?}
One possibility is that the disks cool quickly, such that the dust temperature perturbation along the spiral wake is small, and does not enhance the spiral's intensity contrast (\citealt{speedie22-alma-dustspirals}, see also \citealt{miranda-rafikov20-cooling-basictheory, zhang-zhu20-radiative-cooling}). If that is the case, then we are mainly only probing the spiral \textit{surface density} perturbation, which may be washed out at Band 6/7 wavelengths if the optical depth is sufficiently high. Follow-up at longer observing wavelengths may rule this possibility more or less likely. 
Additionally, in HD 163296 and HD 143006, the planet candidates are embedded in deep gaps and surrounded on either side by narrow rings. Only a small portion of the HD 163296 \#2 and P94 spirals have the opportunity to be expressed upon the rings before they encounter the D48 gap or the outer edge of the continuum disk (Col. 4, Fig \ref{fig:sec3:non-detections-2}).

\subsection{Inconclusive: HD 97048} \label{subsec:inconclusive}

We find strong and large-scale continuum residuals for the two assumed geometries (Cols. 7 \& 8, Table \ref{tab:disks}) for HD 97048 (Fig. \ref{fig:sec3:inconclusive}). Significant positive residuals in the inner disk align with the prediction for the inner spiral under both geometries, and the residuals show a portion of the outer spiral under the \citet{ginski16-hd97048-sphere} geometry. It is unclear whether these matches support the planet hypothesis or are coincidental, because (a) the quality of the match depends on the geometry assumed, and (b) no matter what geometry we assume, there are significant residuals. Considering the possibility that these strong large-scale residuals indicate that an axisymmetric background model is not a good model, we attempt to find spiral residuals by a method that does not assume axisymmetry (Appendix Fig. \ref{fig:app:high-pass}), but come up empty-handed. We thus classify this case inconclusive.

\subsection{Elias 27, IM Lup \& WaOph 6} \label{subsec:detections} 

\begin{figure*}
\begin{center}
\includegraphics[width=18cm]{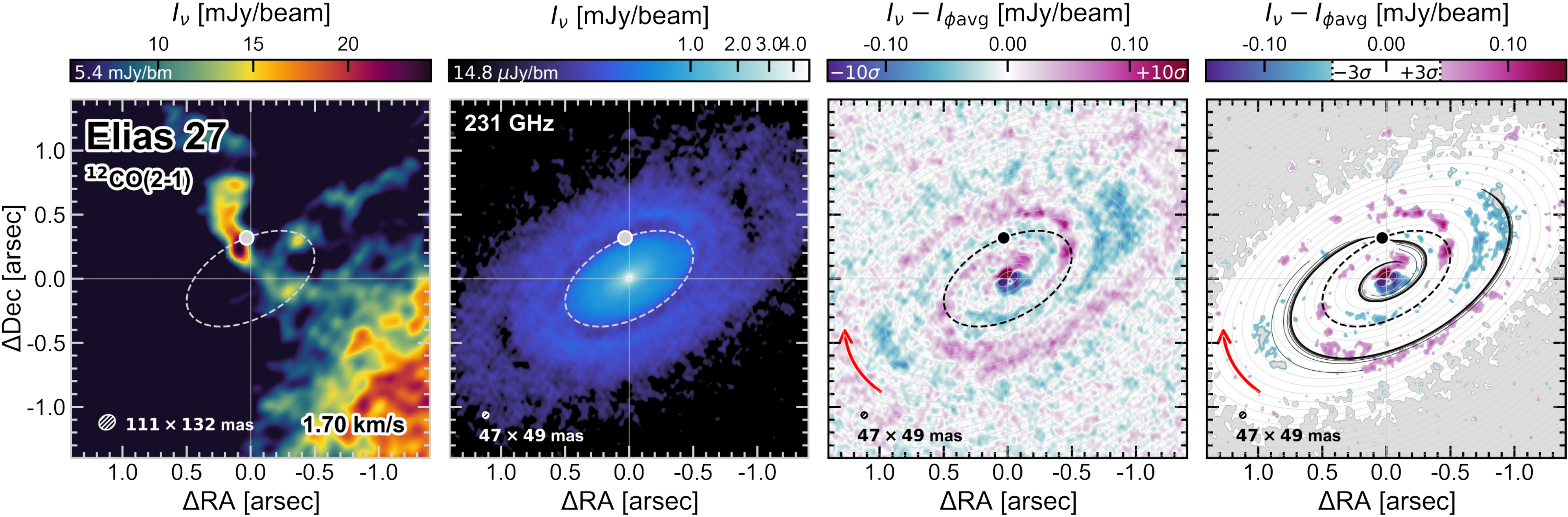}
\includegraphics[width=18cm]{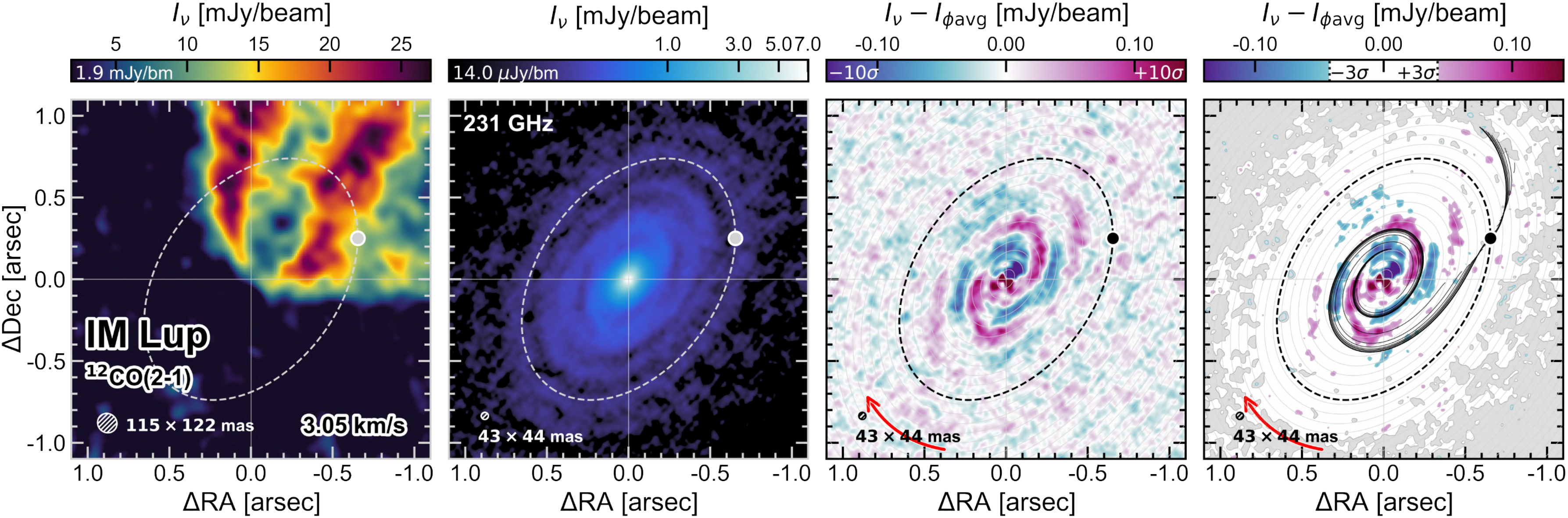}
\includegraphics[width=18cm]{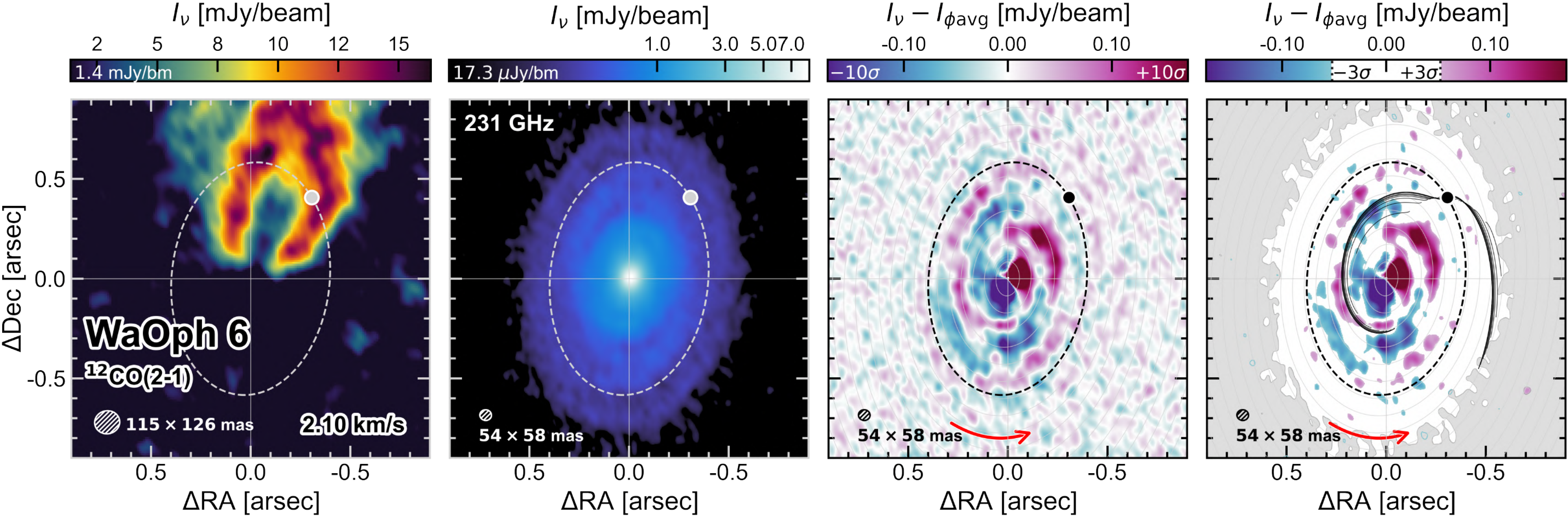}
\end{center}
\caption{\textbf{Detections of dust spirals offset from the predicted planet location:} Elias 27, IM Lup and WaOph 6. The continuum spirals in these disks were first reported in \new{\citet{perez16-elias27} and} \citet{huang18-dsharp3-spirals}.
\label{fig:sec3:detections}}
\end{figure*}

\begin{figure*}
\begin{center}
\includegraphics[width=18cm]{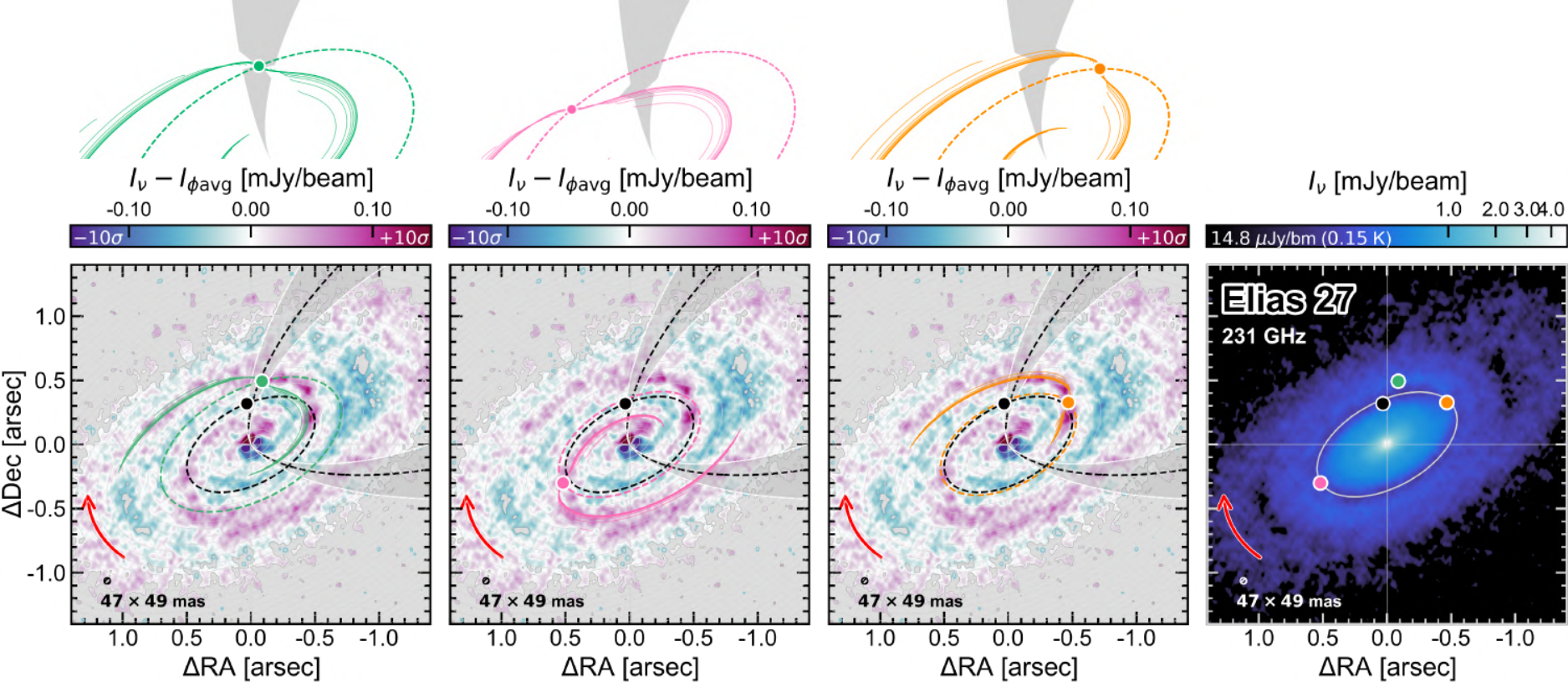}
\includegraphics[width=18cm]{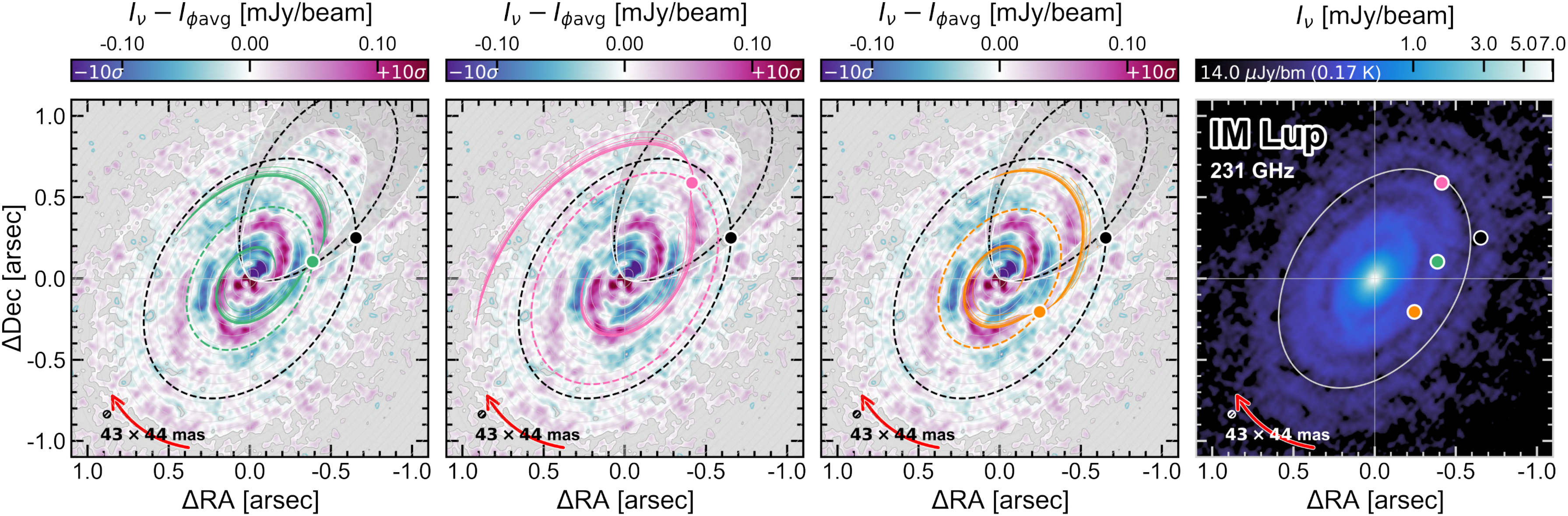}
\includegraphics[width=18cm]{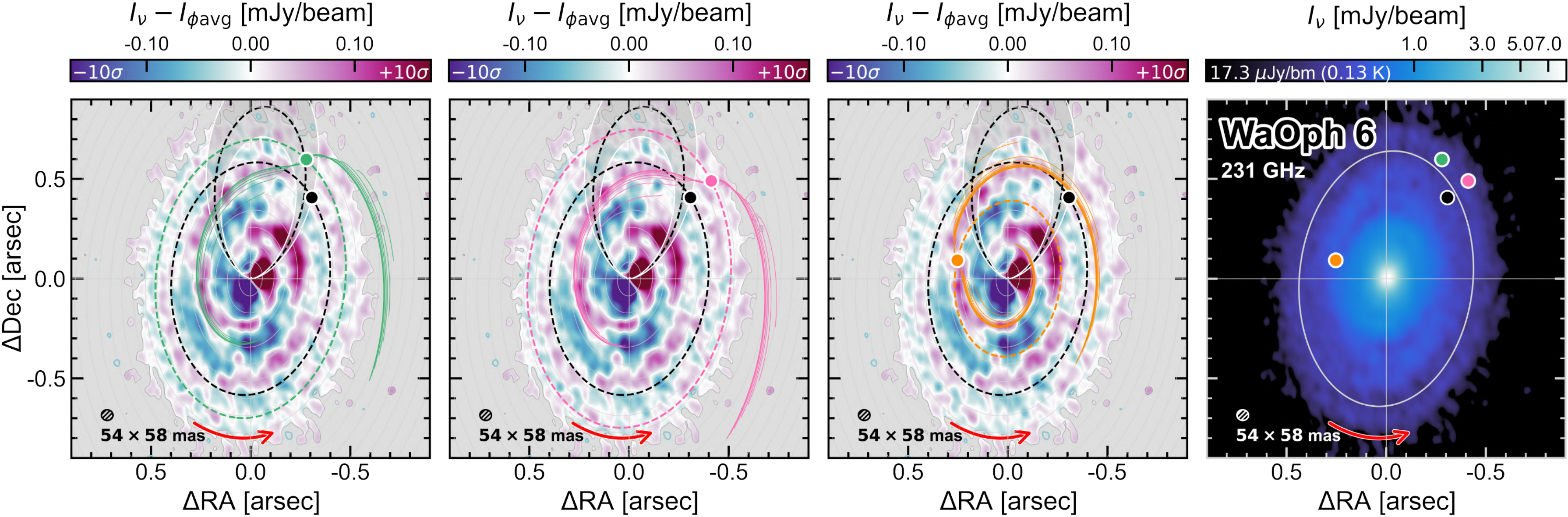}
\end{center}
\caption{\textit{Columns 1-3:} Alternative planet locations that achieve a better match to the observed continuum spiral residuals in Elias 27, IM Lup and WaOph 6, under three different restrictions (represented by the schematic at the top of each column; see \S\ref{subsec:detections} for details). The dashed black isovelocity contour is the velocity of the channel in which the kink is most prominently detected, and shaded grey regions demarcate $\pm 0.5 \times$ the channel width. In all columns, the black dot marks the deprojected kink location (shown in previous figures as representing the reported planet).\\
\textit{Column 4:} Comparison between deprojected kink location and the alternative planet locations. The solid-line white ellipse is the DSHARP dust gap associated with the reported planet location (Col. 10 of Table \ref{tab:kinks}). 
\label{fig:sec4:where-is-the-planet}}
\end{figure*}

Of the 3 detections of dust spirals (Fig. \ref{fig:sec3:detections}), 2 correspond to velocity kinks that were classified as ``firm'' detections \citepalias[Elias 27, IM Lup;][]{pinte20-dsharpkinks}, and 1 corresponds to a ``tentative'' kink detection \citepalias[WaOph 6;][]{pinte20-dsharpkinks}. In all 3 cases, we see two spiral arms in the continuum residual maps, echoing \new{\citet{perez16-elias27} and} \citet{huang18-dsharp3-spirals}. 

\new{Comparing the continuum residuals to the predicted spiral trajectories in Fig. \ref{fig:sec3:detections}, we find that the locations of the detected dust spirals in these 3 disks does not match with where we expect them to lie, given the predicted planet locations.} 

By comparing our estimation of $(H/r)_{\rm p}$ and the reported estimates of the embedded planet masses (Cols. 7 \& 8 of Table \ref{tab:planets}) to Figure 3 of \citet{bae+zhu18-spirals2}, we see that the Elias 27, IM Lup and WaOph 6 planet candidates lie in a region of parameter space where we expect to see both a primary and secondary spiral in the inner disk. \new{This may support the hypothesis that the observed two-armed spirals are planet-driven.}

However, the two arms in these 3 disks are roughly symmetric, and simulations have shown that a planet-to-star mass ratio of $q\sim 0.01$ (or larger)
is required to make symmetric inner primary and secondary spiral arms \cite[Fig. 3,][]{fung-dong15-inferring-planet-mass-spirals}. This is a point of mild tension with the 
masses inferred from the velocity kink amplitude ($q \in [0.001, 0.005]$ in these 3 cases). \new{Gravitational instability may be 
a better
explanation for symmetric two-armed spirals (e.g., for Elias 27, see \citealt{meru17-elias27,tomida17-elias27-GI, paneque-carreno21-elias27-GI}, and for such spirals in other disks, \citealt{2015ApJ...812L..32D}).} 

\new{As an alternative possibility that maintains the planetary-origin hypothesis,} we contemplate the method used to determine the predicted planet locations (\S\ref{subsec:sample}), which involves deprojecting the visually-identified kink center location from the estimated emission surface directly onto the midplane. Since the $\CO$ emission surface is expected to be a few scale heights above the midplane \new{\new{\citep[e.g., ][]{pinte2018-altitude-of-CO, law21-maps4-emissionsurfaces, paneque-carreno22-vertical-stratification}}}, there may be room for error in translation. For example, \citet{zhu15-3dstructure-spiralshocks} showed with 3D hydrodynamical simulations that spiral wakes are not perpendicular to the midplane, and instead curl towards the star at the disk surface. Vertical temperature gradients can introduce further complications, changing a spiral's pitch angle and misaligning the surface wakes from the midplane wakes \new{\citep{juhasz-rosotti18-pluto, rosotti20-hd100453}}. 
\new{While \citet{calcino22-hd163296} had success in matching $\CO$ emission surface kinks around the full disk azimuth to the predicted spiral wake of HD 163296 \#1 (P261), their simulations and analytic models assumed no vertical dependence in the velocity perturbations, and the location of this planet (which lies outside the continuum) has not been confirmed in midplane tracers.}
\new{It's therefore conceivable that the method for pinpointing the planet may need to encompass vertical} effects. 
Motivated by the possibility of leveraging the location of the midplane continuum spirals to inform the development of such a method, we \new{assume the spirals in Elias 27, IM Lup and WaOph 6 are planet-driven,} and explore alternative planet locations. 

\subsubsection{Considering alternative planet locations}  \label{subsubsec:alternative-locations}

In the following, we present 3 alternative planet locations (shown left to right in Fig. \ref{fig:sec4:where-is-the-planet}) in each of Elias 27, IM Lup and WaOph 6, under a set of 3 informative and gradually loosening restrictions. In all cases, the restrictions are based on \textit{midplane} information: the dust continuum residuals, the deprojected kink location, midplane isovelocity contours, and the 2D velocity kink theory of \citet{bollati21-theory-of-kinks-2d}. \new{We calculate the midplane isovelocity contours using the 2D Keplerian velocity field $v_0(R, \phi) = v_{\rm Kep}(R) \, \sin{(i)} \, \cos{(\phi)} + v_{\rm LSR}$, where $i$ is the disk inclination, $R$ and $\phi$ are the disk frame coordinates ($\phi$ measured from the redshifted disk major axis), and $v_{\rm LSR}$ is the systemic velocity. For $v_{\rm LSR}$ values see caption of Table \ref{tab:possible-planets}.}

In the 1st column of Fig. \ref{fig:sec4:where-is-the-planet}, 
we shift the planet location to get a better alignment with the detected dust spirals, under the restriction that the planet cannot lie outside the midplane area of the channel in which the velocity kink is most prominently detected. The motivation for this restriction is the idea that the velocity kink amplitude should be strongest close to the planet \new{\citep[e.g.,][]{bollati21-theory-of-kinks-2d, calcino22-hd163296}}. 
We represent this area in grey, which spans a half channel width on either side of the velocity of the kink channel\footnote{In the case of Elias 27, we infer from Table 2 and Fig. 1 of \citetalias{pinte20-dsharpkinks} that the kink is also detected in the two adjacent channels, though strong cloud contamination is present. In the case of Elias 27 and WaOph 6, it is unclear from their Table 2 whether the kink is detected in more than one channel. We thus opt to only consider the single channel.} (Col. 6, Table \ref{tab:kinks}) in order to incorporate the spatial ``uncertainty'' introduced by the spectral resolution of the CO data.  We are able to achieve more satisfactory alignments, but find that the necessary shift in radial and azimuthal position places the planets outside their DSHARP dust gap (white solid ellipse in the 4th column).

At the top of Fig. \ref{fig:sec4:where-is-the-planet}, we show a midplane schematic of how we may expect the planet location to affect the emission morphology in a given channel \citep[a logic-extension of the results from][]{bollati21-theory-of-kinks-2d}: if the inner spiral wake shifts emission to lower velocity channels, and the outer wake shifts emission to higher velocity channels, then the channel centered on the planet may be left with an absence emission at the planet's location. The kink (specifically, emission present in a channel that is spatially offset from the rest) may then instead be most prominent in a channel where it coincides with the inner or outer spiral wake.

Thus, in the 2nd and 3rd columns of Fig. \ref{fig:sec4:where-is-the-planet}, we again shift the planet location to get a better alignment with the detected dust spirals, 
but this time while maintaining that the reported kink is probing a portion of the planet's inner spiral arm (2nd column) or outer spiral arm (3rd column), with the planet being as close to the deprojected kink location as possible. 
Under these two restrictions, we find some qualitative improvement in the match to the detected dust spiral, and in some cases (inner and outer wake scenario for Elias 27, inner wake scenario for IM Lup) we find that the resulting planet location lies inside the DSHARP dust gap. Important to note is that the \textit{observed} midplane dust spiral residuals do not perfectly intersect with the deprojected kink location (and so the inner and outer wakes of our planet locations do not achieve perfect intersection either), suggesting a possible disjunction between midplane spirals and their expression on the disk surface.

We consider the 3 planet locations for each disk in Fig. \ref{fig:sec4:where-is-the-planet} to be \textit{possible} locations, in the sense that they plausibly satisfy the continuum spiral residuals. The main caveat is that we have not quantitatively assessed the agreement between the theoretical spiral \new{trajectories} and the continuum residuals, and obtained the planet locations by visual inspection / trial and error. We provide the locations in Table \ref{tab:possible-planets}, and note that in some cases the planet in the midplane lies far from the deprojected velocity kink in $\CO$ surface emission. As mentioned above, our determination of these planet locations was done using midplane-based information, without consideration for any surface velocity evidence associated with the new locations, and how the planets can reproduce the strength of the detected kink signals in a distant channel is unclear. It may not be the case that the channel in which the kink is intrinsically most prominent has been correctly identified in Elias 27 and WaOph 6, though, as the $\CO$ channel maps of these 2 disks suffer cloud contamination. \new{This applies to almost the entire redshifted (south) half of Elias 27, and a large portion of WaOph 6 from the disk minor axis toward the blueshifted (north) side (see note $a$ in Table \ref{tab:kinks} for affected velocities).}

Our results emphasize the need for more theoretical and simulation work to understand the expected morphology of a planet-driven velocity kink, how the planet's spiral manifests at different heights in the disk, how the strength of the kink signal should vary with channel, and how we can use that information to successfully pinpoint the planet.

\section{Summary} \label{sec:summary}

\begin{enumerate}
    \item \new{Despite the sufficiently high planet masses inferred from the reported velocity kink amplitudes, we are unsuccessful in detecting any dust spirals associated with 6 of the 10 velocity kink planet candidates reported to date whose orbits lie within the continuum disk, using current continuum observations (Figs. \ref{fig:sec3:non-detections-1} \& \ref{fig:sec3:non-detections-2}). We interpret this to mean that the full planet-finding potential of the velocity kink method may not be exemplified by this specific set of candidates. More kink detection efforts, including better quantification of the kink signal robustness and assessment for a planet-driven morphology, are needed.} 
    \item \new{Our search for dust spirals in the HD 97048 disk is inconclusive (Fig. \ref{fig:sec3:inconclusive}). Observations with higher resolution and/or better sensitivity are needed to renew the search.}
    \item \new{In the remaining 3 disks in our sample (Elias 27, IM Lup and WaOph 6), we re-detect clear and coherent dust spirals in the continuum residuals \citep{perez16-elias27, huang18-dsharp3-spirals}, but find that they do not align with the theoretical spiral trajectory originating at the candidate planet's reported location (Fig. \ref{fig:sec3:detections}). If these spirals are planet-driven, then this spatial offset may indicate that the method used to pinpoint the planet location from the kink detection in these disks (\S\ref{subsec:sample}) is incomplete; a more successful method may need to encompass how a midplane spiral can be ``morphed'' during its upward propagation to be expressed on the disk surface (\S\ref{subsec:detections}). We provide alternative midplane planet locations that are plausible from the dust spiral's perspective for these 3 planet candidates in Fig. \ref{fig:sec4:where-is-the-planet} and Table \ref{tab:possible-planets}, which in some cases are far from the reported velocity kink (\S\ref{subsubsec:alternative-locations}).}
\end{enumerate}


\vspace{5mm}

\new{We thank the anonymous referee for their thoughtful and constructive questions and suggestions.}
J.S. thanks Richard Booth, Cathie Clarke, Giovanni Rosotti, Richard Alexander, Richard Nelson, Brodie Norfolk, Rebecca Nealon, Sahl Rowther, Guilia Ballabio, Simon Casassus, Sebasti\'an P\'erez and Philipp Weber for insightful discussions that helped shape this work, \nuwe{and we thank Daniel Price for comments on the manuscript}. 
J.S. also thanks the curator of the Catalog of Circumstellar Disks (\href{https://www.circumstellardisks.org/index.php}{www.circumstellardisks.org}).
R.D. and J.S. are supported by the Natural Sciences and Engineering Research Council of Canada (NSERC) and the Alfred P. Sloan Foundation.

We are grateful to Christophe Pinte and DSHARP Collaboration for making their data publicly available. This paper makes use of the following ALMA data: ADS/JAO.ALMA \#2016.1.00484.L, ADS/JAO.ALMA \#2016.1.00825.S. ALMA is a partnership of ESO (representing its member states), NSF (USA) and NINS (Japan), together with NRC (Canada), MOST and ASIAA (Taiwan), and KASI (Republic of Korea), in cooperation with the Republic of Chile. The Joint ALMA Observatory is operated by ESO, AUI/NRAO and NAOJ. The National Radio Astronomy Observatory is a facility of the National Science Foundation operated under cooperative agreement by Associated Universities, Inc.
This work has made use of data from the European Space Agency (ESA) mission
{\it Gaia} (\url{https://www.cosmos.esa.int/gaia}), processed by the {\it Gaia}
Data Processing and Analysis Consortium (DPAC,
\url{https://www.cosmos.esa.int/web/gaia/dpac/consortium}). Funding for the DPAC
has been provided by national institutions, in particular the institutions
participating in the {\it Gaia} Multilateral Agreement.


\vspace{5mm}
\facility{ALMA.}

\software{\texttt{astropy} \citep{software-astropy1, software-astropy2}, 
          \texttt{cmasher} \citep{software-cmasher}, 
          \texttt{disksurf} \citep{software-disksurf}, 
          \texttt{gofish} \citep{software-gofish}, 
          \texttt{matplotlib} \citep{software-matplotlib}, 
          \texttt{numpy} \citep{software-numpy}, 
          \texttt{pandas} \citep{software-pandas}, 
          \texttt{scipy} \citep{software-scipy}. 
          }

\appendix

\restartappendixnumbering 

\section{Tables} \label{app:tables}

Tables \ref{tab:planets}, \ref{tab:observations}, \ref{tab:disks} and \ref{tab:possible-planets}.

\begin{deluxetable*}{cccccccc}
\tablenum{A1}
\tablecaption{Inferred midplane locations of planets detected by a velocity kink inside the continuum. \label{tab:planets}}
\tabletypesize{\scriptsize}
\tablehead{
\colhead{Disk} & \colhead{Ref.} & \multicolumn2c{\underline{Planet Sky Coordinates}} & \multicolumn2c{\underline{Planet Disk Frame Coordinates}} &\colhead{$(H/r)_{\rm p}$} & \colhead{$\Mp$ } \\ 
\colhead{ } & \colhead{ } & \colhead{$\rpsky$} & \colhead{$\PApsky$} & \colhead{$\rpdisk$} & \colhead{$\PApdisk$} & \colhead{ } & \colhead{ }    \\
\colhead{ } & \colhead{ } & \colhead{($\arcsec$)} & \colhead{(deg)} & \colhead{(au)} & \colhead{(deg)}  & \colhead{ }  & \colhead{($\Mth$) } 
}
\decimalcolnumbers
\startdata
Elias 27        & \citetalias{pinte20-dsharpkinks}     & 0.32 $\pm$ ...   & 6    $\pm$ ... & 60  $\pm$ ... & -103   $\pm$ ... & 0.087 & $3.0-8.9$  \\
HD 143006       & \citetalias{pinte20-dsharpkinks}     & 0.14 $\pm$ ...   & 107  $\pm$ ... & 24  $\pm$ ... & -61  $\pm$ ...   & 0.043 & $6.5-19.6$  \\
HD 163296 (P94) & \citetalias{izquierdo22-hd163296}    & 0.77 $\pm$ 0.05  & -8   $\pm$ 3   & 94  $\pm$ 6   & 50     $\pm$ 3   & 0.069 & $1.4-4.3$  \\
HD 163296 (\#2) & \citetalias{pinte20-dsharpkinks}     & 0.67 $\pm$ ...   & 93   $\pm$ ... & 82  $\pm$ ... & 129   $\pm$ ...  & 0.066 & $1.6-4.8$  \\
IM Lup          & \citetalias{pinte20-dsharpkinks}     & 0.70 $\pm$ ...   & -69  $\pm$ ... & 127 $\pm$ ... & 136  $\pm$ ...   & 0.089 & $1.5-4.6$  \\
DoAr 25         & \citetalias{pinte20-dsharpkinks}     & 0.44 $\pm$ ...   & -36  $\pm$ ... & 101 $\pm$ ... & 60   $\pm$ ...   & 0.071 & $2.8-8.3$  \\
GW Lup          & \citetalias{pinte20-dsharpkinks}     & 0.42 $\pm$ ...   & -89  $\pm$ ... & 78  $\pm$ ... & -119  $\pm$ ...  & 0.084 & $3.5-10.4$  \\
Sz 129          & \citetalias{pinte20-dsharpkinks}     & 0.29 $\pm$ ...   & -148 $\pm$ ... & 53  $\pm$ ... & 63     $\pm$ ... & 0.059 & $5.6-16.8$  \\
WaOph 6         & \citetalias{pinte20-dsharpkinks}     & 0.51 $\pm$ ...   & -37  $\pm$ ... & 72  $\pm$ ... & 138    $\pm$ ... & 0.089 & $2.0-6.0$  \\
HD 97048        & \citetalias{pinte19-hd97048}         & 0.45 $\pm$ 0.10  & -55  $\pm$ 10  & 109 $\pm$ 24  & -66    $\pm$ 10  & 0.073 & $2.1-3.1$  \\
\enddata
\tablecomments{Column descriptions:
(1) Name of disk. Name of kink or planet candidate in brackets, if applicable.
(2) Reporting paper, as in Table \ref{tab:kinks}.
(3-4) Coordinates of the planet as seen on the sky: Radial separation from the star ($\rpsky$), and position angle measured east of north ($\PApsky$). The `...' indicates where authors gave no indication of uncertainty. Note that \citetalias{pinte20-dsharpkinks} (their Table 1) provides $\PApsky$ measured \textit{west} of north.
(5-6) Coordinates of the planet in the disk frame: Radius in the deprojected midplane ($\rpdisk$), and polar angle measured anti-clockwise from the disk's redshifted major axis ($\PApdisk$). Disk frame coordinates were calculated by this work, with the exception of row 3 \citepalias[P94,][]{izquierdo22-hd163296}, in which case we calculated the sky frame coordinates. Values of $d$ used for arcsec$\leftrightarrow$au are in Table \ref{tab:disks}.
(6) Aspect ratio $(H/r)$ evaluated at $\rpdisk$, calculated by this work using Eqn. \ref{eqn:Tmid-dullemond18} and $\Lstar$ in Table \ref{tab:disks}.
(7) Mass estimate of the planet in units of thermal mass $\Mth$, calculated by this work using Column 6, $\Mstar$ in Table \ref{tab:disks}, and the $\Mp$ range in units of $\Mjup$ from the reporting paper (Column 11 of Table \ref{tab:kinks}).\\
}
\end{deluxetable*}

\begin{deluxetable}{cccccc}
\tablenum{A2}
\tablecaption{Summary of observations used in this work. \label{tab:observations}}
\tablewidth{0pt}
\tabletypesize{\scriptsize}
\tablehead{
\colhead{Disk}  & \colhead{Origin} & \multicolumn2c{CO} & \multicolumn2c{Continuum}  \\
\colhead{ }  & \colhead{ } & \colhead{rms noise} & \colhead{$\theta_{\rm beam}$} & \colhead{rms noise} & \colhead{$\theta_{\rm beam}$}  \\
\colhead{ }  & \colhead{ } & \colhead{(mJy/bm)} & \colhead{(mas)} & \colhead{($\mu$Jy/bm)} & \colhead{(mas)}
}
\decimalcolnumbers
\startdata
Elias 27  & DDR    & 1.6 & $111 \times 132$  & 14.8 & $47 \times 49$   \\
HD 143006 & DDR    & 1.0 & $49 \times 66$  & 10.7 & $36 \times 53$  \\
HD 163296 & DDR    & 0.8 & $95 \times 104$  & 22.6 & $38 \times 48$   \\
IM Lup    & DDR    & 1.9 & $115 \times 122$  & 14.0 & $43 \times 44$   \\
DoAr 25   & DDR    & 1.3 & $78 \times 101$  & 12.9 & $22 \times 41$   \\
GW Lup    & DDR    & 1.3 & $81 \times 109$  & 15.1 & $43 \times 45$   \\
Sz 129    & DDR    & 1.0 & $83 \times 110$  & 15.5 & $31 \times 44$   \\
WaOph 6   & DDR    & 1.4 & $115 \times 126$  & 17.3 & $54 \times 58$   \\
HD 97048  & FS     & 3.7 & $70 \times 110$  & 76.9 & $66 \times 101$   \\
\enddata
\tablecomments{Column descriptions:
(1) Name of source.
(2) DDR: \href{https://almascience.eso.org/almadata/lp/DSHARP/}{DSHARP Data Repository} (Program ID: 2016.1.00484.L), FS: \href{https://figshare.com/articles/dataset/HD_97048_ALMA_B7_continuum_13CO/8266988}{Figshare} (Program ID: 2016.1.00825.S).
(3) Measured rms noise in the cube. 
(4) Synthesized beam FWHM of the cube.
(5) Measured rms noise in the image. 
The values in Column 3 \& 5 are almost identical to those of \citet{andrews18-dsharp1} (Table 4 \& 5) and \citet{pinte19-hd97048}.
(6) Synthesized beam FWHM of the continuum image.
}
\end{deluxetable}


\begin{deluxetable*}{cccccccccc}
\tablenum{A3}
\tablecaption{Stellar properties and disk geometries. \label{tab:disks}}
\tablehead{
\colhead{Disk} & \colhead{$d$} & \colhead{$\Mstar$} & \colhead{$\Lstar$} & \multicolumn5c{\underline{\hspace{3cm} Disk geometry\hspace{3cm}}} & \colhead{Rotation}\\
\colhead{ } & \colhead{ } & \colhead{ } & \colhead{ } & \colhead{$\xo$} & \colhead{$\yo$} & \colhead{$i$} & \colhead{PA} & \colhead{Method} & \colhead{ } \\
\colhead{ } & \colhead{(pc)} & \colhead{($\Msun$)} & \colhead{($\Lsun$)} & \colhead{(mas)} & \colhead{(mas)} & \colhead{(deg)} & \colhead{(deg)} & \colhead{ } & \colhead{ } \\
\colhead{(1)} & \colhead{(2)} & \colhead{(3)} & \colhead{(4)} & \colhead{(5)} & \colhead{(6)} & \colhead{(7)} & \colhead{(8)} & \colhead{(9)} & \colhead{(10)}  
} 
\startdata
Elias 27  & 110.1   $\pm$ 10.3  & 0.49$_{-0.11}^{+0.20}$  & 0.91$_{-0.37}^{+0.64}$  & -5   $\pm$ 5     & -8   $\pm$ 3     & 56.2 $\pm$ 0.8    & 118.8 $\pm$ 0.7      & E   & CW   \\
HD 143006 & 167.3   $\pm$ 0.5   & 1.78$_{-0.30}^{+0.22}$  & 3.8$_{-1.1}^{+1.6}$   & -6   $\pm \sim$2 & 23   $\pm \sim$2 & 16   $\pm \sim$2  & 167   $\pm \sim$2    & FRANK   & CW?   \\
HD 163296 & 101.0   $\pm$ 0.4   & 2.04$_{-0.13}^{+0.25}$  & 17.0$_{-8.5}^{+17}$  & -3.5 $\pm \sim$2 & 4    $\pm \sim$2 & 47   $\pm \sim$2  & 313   $\pm \sim$2    & FRANK   & CW   \\
IM Lup    & 155.8   $\pm$ 0.5   & 0.89$_{-0.23}^{+0.21}$  & 2.6$_{-0.95}^{+1.5}$   & -1.5 $\pm$ 2     & 1    $\pm$ 2     & 47.5 $\pm$ 0.3    & 144.5 $\pm$ 0.5      & E   & CW    \\
DoAr 25   & 138.2   $\pm$ 0.8   & 0.95$_{-0.33}^{+0.10}$  & 1.0$_{-0.35}^{+0.56}$   & 38   $\pm$ 2     & -494 $\pm$ 2     & 67.4 $\pm$ 0.2    & 290.6 $\pm$ 0.2      & E   & CW    \\
GW Lup    & 155.2   $\pm$ 0.4   & 0.46$_{-0.15}^{+0.12}$  & 0.33$_{-0.12}^{+0.19}$  & 0.5  $\pm \sim$2 & -0.5 $\pm \sim$2 & 39   $\pm \sim$2  & 37    $\pm \sim$2    & FRANK   & ACW?   \\
Sz 129    & 160.1   $\pm$ 0.4   & 0.83$_{-0.24}^{+0.06}$  & 0.44$_{-0.16}^{+0.26}$  & 5    $\pm \sim$2 & 6    $\pm \sim$2 & 32   $\pm \sim$2  & 153   $\pm \sim$2    & FRANK   & ?   \\
WaOph 6   & 122.5   $\pm$ 0.4   & 0.68$_{-0.13}^{+0.32}$  & 2.9$_{-1.0}^{+1.7}$   & -244 $\pm$ 3     & -361 $\pm$ 3     & 47.3 $\pm$ 0.7    & 174.2 $\pm$ 0.8      & E   & ACW    \\
HD 97048  & 184.4   $\pm$ 0.8   & 2.36 $\pm$ 0.19         & 36.6 $\pm$ 20.03     & 0                &  0               & 45.3 $\pm$ 2.55   & 2.84  $\pm$ 2.55     & Velocity Field   & ACW \\
          &                     &                         &                      & 0                &  0               & 39.9 $\pm$ 1.8   & 2.8  $\pm$ 1.6     & Scattered Light   &  \\
\enddata
\tablecomments{Column descriptions: 
(1) Name of source. 
(2) Distance to the source, from \textit{Gaia} DR3 \citep{2016-gaia-collab-mission, gaia22-vallenari} as $d=1/\varpi$.
(3) Stellar mass.
(4) Stellar luminosity. Values for $\Lstar$, and $\Mstar$ are from \citet{andrews18-dsharp1} except HD 97048, in which case $\Lstar$ and $\Mstar$ are from \citet{bohn22-misalignments}.
(5) R.A. offset of disk center from phase center (in the datasets we use; see Table \ref{tab:observations}).
(6) Decl. offset of disk center from phase center.
(7) Disk inclination.
(8) Disk position angle, measured anti-clockwise (i.e., east of north) to the red-shifted major axis.
(9) Method used to estimate disk P.A., inclination, and offset from phase center in the work from which we source the values: ``E'' indicates that the values were derived by fitting ellipses to continuum annular substructures \citep[Table 2,][]{huang18-dsharp2}; ``FRANK'' indicates the \texttt{frank} residual appearance method of \citet{andrews21-frank-dsharp} (their Table 2); ``Velocity Field'' indicates the results of fitting a Keplerian disk model to the velocity field from CO line data cubes \citep[Table C.1,][]{bohn22-misalignments} with \texttt{eddy} \citep{teague19-eddy}; and ``Scattered Light'' indicates the results of fitting ellipses to gaps and rings observed in near-IR scattered light by SPHERE \citep[Table 1,][]{ginski16-hd97048-sphere}.
(10) Direction in which the disk rotates. ``CW'' means clockwise (west of north) and ``ACW'' means anti-clockwise. Rotation direction for Elias 27 and WaOph 6 were taken from \citet{huang18-dsharp3-spirals}; the remaining disk rotations were determined by this work based on scattered light observations in the following works: 
Elias 27 \citep{huang18-dsharp3-spirals}, 
HD 143006 \citep{benisty18-hd143006-SL, perezL18-hd143006}, 
HD 163296 \citep{monnier17-hd163296-SL, muroarena18-hd163296-SL}, 
IM Lup \citep{avenhaus18-sphere, huang18-dsharp3-spirals}, 
DoAr 25 \citep{andrews08-doar25, garufi20-DARTTS}, 
GW Lup \citep{garufi22-sphere}, 
Sz 129 (none to our knowledge), 
WaOph 6 \citep{huang18-dsharp3-spirals}, 
HD 97048 \citep{ginski16-hd97048-sphere}. 
``?'' indicates cases where the near/far side is uncertain or unknown in scattered light images in the literature to date (note the low inclination of those 3 disks).
}
\end{deluxetable*}

\begin{deluxetable}{ccccccc}
\tablenum{A4}
\tablecaption{Coordinates of the alternative planet locations presented in Fig. \ref{fig:sec4:where-is-the-planet}. \label{tab:possible-planets}}
\tabletypesize{\scriptsize}
\tablewidth{0pt}
\tablehead{
\colhead{Disk} & \colhead{Col.} & \multicolumn2c{\underline{Sky Coordinates}} & \multicolumn2c{\underline{Disk Frame Coords.}}& \colhead{$v_{\rm p}$}\\
\colhead{ } & \colhead{ } & \colhead{$\rpsky$} & \colhead{$\PApsky$} & \colhead{$\rpdisk$} & \colhead{$\PApdisk$} & \colhead{}    \\
\colhead{ } & \colhead{ } & \colhead{($\arcsec$)} & \colhead{(deg)} & \colhead{(au)} & \colhead{(deg)} & \colhead{(km/s)} 
}
\decimalcolnumbers
\startdata
Elias 27 & 1      & 0.50    & -10   & 85     & -114   & 1.65   \\
         & 2      & 0.60    & 120   & 66     & 2      & 4.48   \\
         & 3      & 0.57    & -55   & 64     & 191    & 0.31   \\
IM Lup   & 1      & 0.40    & -75   & 76     & 129    & 2.99 \\
         & 2      & 0.72    & -35   & 112    & -179   & 2.51 \\
         & 3      & 0.32    & -130  & 74     & 87     & 4.57 \\
WaOph 6 & 1      & 0.66    & -25   & 86     & 153    & 2.13 \\
         & 2      & 0.64    & -40   & 92     & 135    & 2.53 \\
         & 3      & 0.27    & 70    & 48     & -100   & 3.40 \\
\enddata
\tablecomments{Column descriptions:
(1) Name of disk. 
(2) Column of Fig. \ref{fig:sec4:where-is-the-planet} showing the planet whose coordinates are given, numbered 1 to 3, left to right.
(3-4) Coordinates of the planet as seen on the sky: Radial separation from the star ($\rpsky$), and position angle measured east of north ($\PApsky$).
(5-6) Coordinates of the planet in the disk frame: Radius in the deprojected midplane ($\rpdisk$), and polar angle measured anti-clockwise from the disk's redshifted major axis ($\PApdisk$). Values of $d$ used for arcsec$\leftrightarrow$au are in Table \ref{tab:disks}.
(7) Velocity coordinate of the planet, relative to Earth. Systemic velocities used to find these values were estimated from the morphology of emission in the channel maps (thus having uncertainty $v_{\rm sys}$, Col. 5 of Table \ref{tab:kinks}), and are: $v_{\rm LSR}= 2.40$ km/s for Elias 27, $v_{\rm LSR}= 4.45$ km/s for IM Lup and $v_{\rm LSR}= 3.85$ km/s for WaOph 6. }
\end{deluxetable}

\vspace{-24mm}

\section{Treatment of Confined Azimuthal Asymmetries} \label{app:disk-geometry}

The strong emission from the confined arc-like features in HD 163296 and HD 143006 will, of course, raise the azimuthal average emission within the radial region that they occupy, therefore making it more difficult to detect any dust spirals above the (overly-positive) average background in that radial region. This is particularly relevant for our search in HD 163296, as the inner spiral wake of both the HD 163296 \#2 and P94 planets would be expressed upon the B67 ring, which is contaminated by the arc-like feature. It is less important for our search in HD 143006, because the candidate planet's spiral wake is tightly wound and unlikely to ``reach'' the radial region occupied by the crescent. 
In HD 163296, we omit from the calculation of the azimuthal average the emission lying within $135^{\circ} \leq \phi \leq 220^{\circ}$ between $R=0.49 \arcsec$ and $R=0.63 \arcsec$, where $R$ and $\phi$ are disk frame coordinates ($\phi$ measured anti-clockwise from the redshifted disk major axis). In HD 143006, we omit the emission lying within $-58^{\circ} \leq \phi \leq 12^{\circ}$ between $R=0.37 \arcsec$ and $R=0.5 \arcsec$. 

\begin{figure*}
\begin{center}
\includegraphics[width=9.15cm]{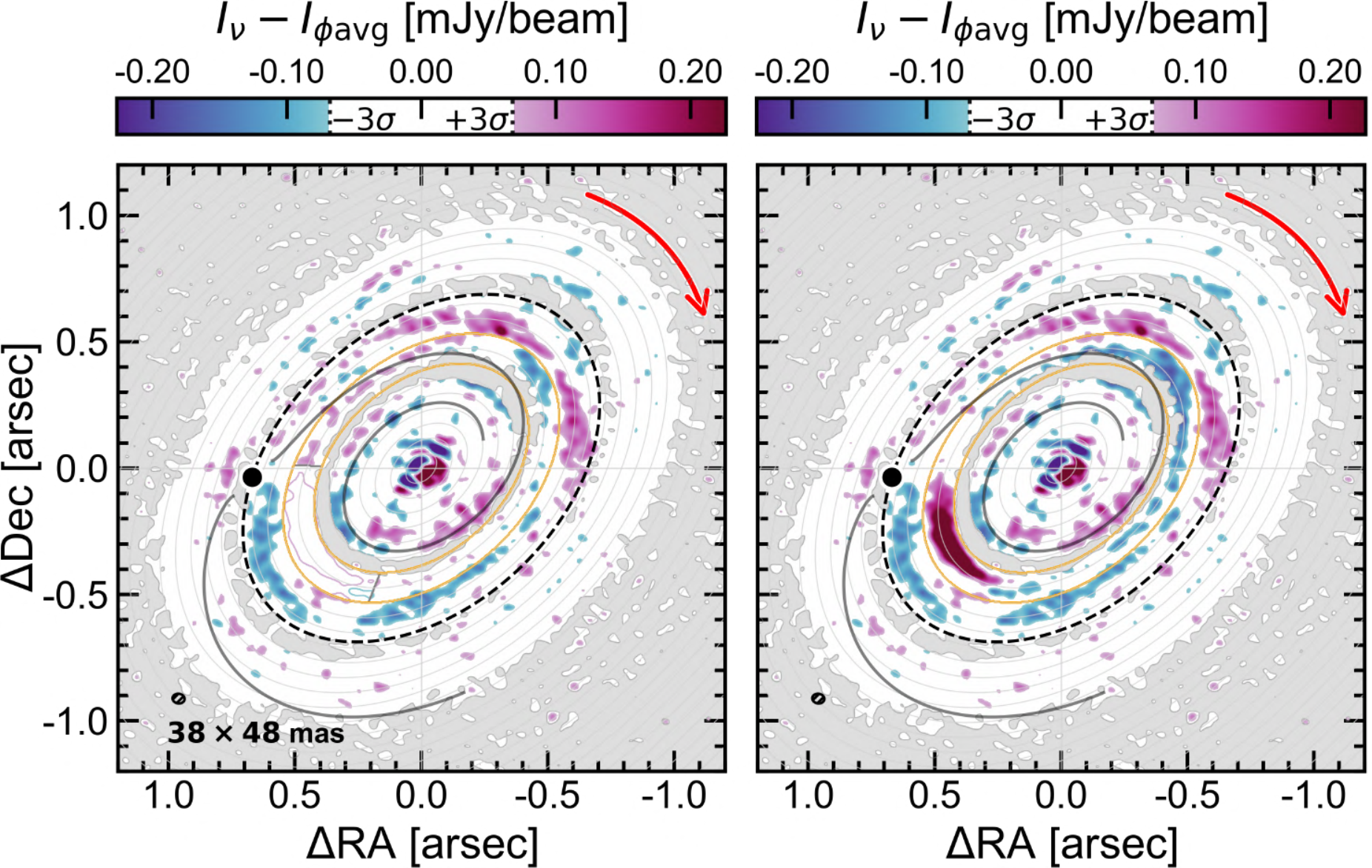}
\includegraphics[width=8.76027777777cm]{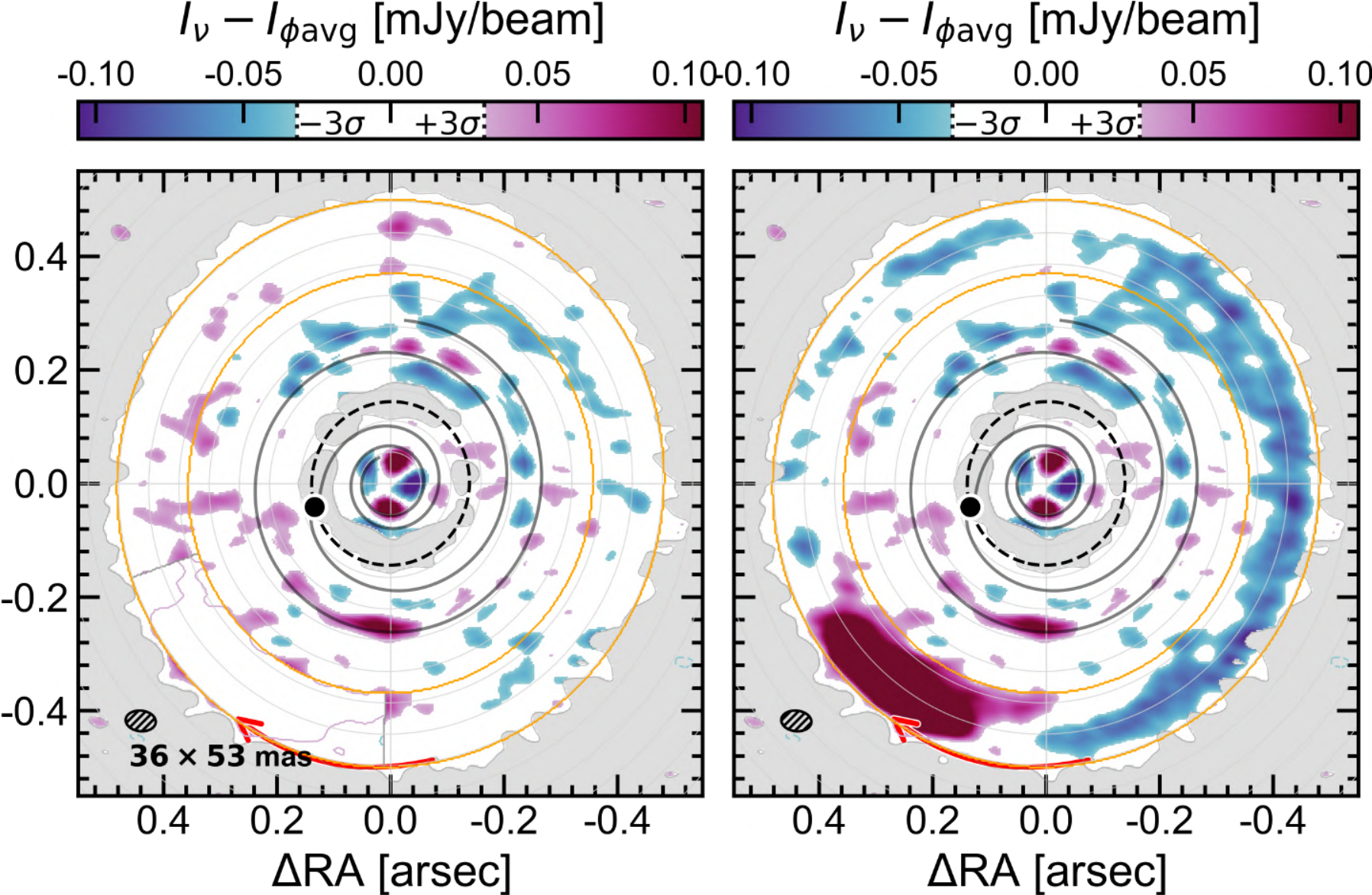}
\end{center}
\caption{\textbf{Treatment of crescents in HD 163296 and HD 143006:} Detected residual substrctures stronger than $3\times$ the continuum rms noise in HD 163296 (left two panels) and HD 143006 (right two panels) when either including or excluding the confined arc-like azimuthal asymmetries. The two orange ellipses guide the eye to the radial region affected (i.e., the residual maps are identical outside of this region).
A $3\sigma$ contour of the arc-like feature is shown under the mask to demonstrate how the mask covers the feature.
\label{fig:app:masking-asymmetries}}
\end{figure*}

Fig. \ref{fig:app:masking-asymmetries} shows the continuum residual maps in HD 163296 and HD 143006 with and without including the confined arc-like features in the azimuthal average. In HD 163296, excluding the arc-like feature modifies the residuals in a way that is more relevant for the HD 163296 \#2 planet candidate than for P94 (hence why it is plotted in Fig. \ref{fig:app:masking-asymmetries} instead of P94) -- some disconnected positive residuals are introduced in the northeast portion of the radial region (where HD 163296 \#2's inner spiral would lie), though they do not appear to be a segment of a spiral. In HD 143006, excluding the arc-like feature removes the strongly negative residuals in the west half of the disk, but doesn't affect the residuals near the planet candidate. 

\section{\new{Re-imaging the continuum visibilities to achieve higher sensitivity}} \label{app:re-imaging}

\begin{figure}
\begin{center}
\includegraphics[width=\columnwidth]{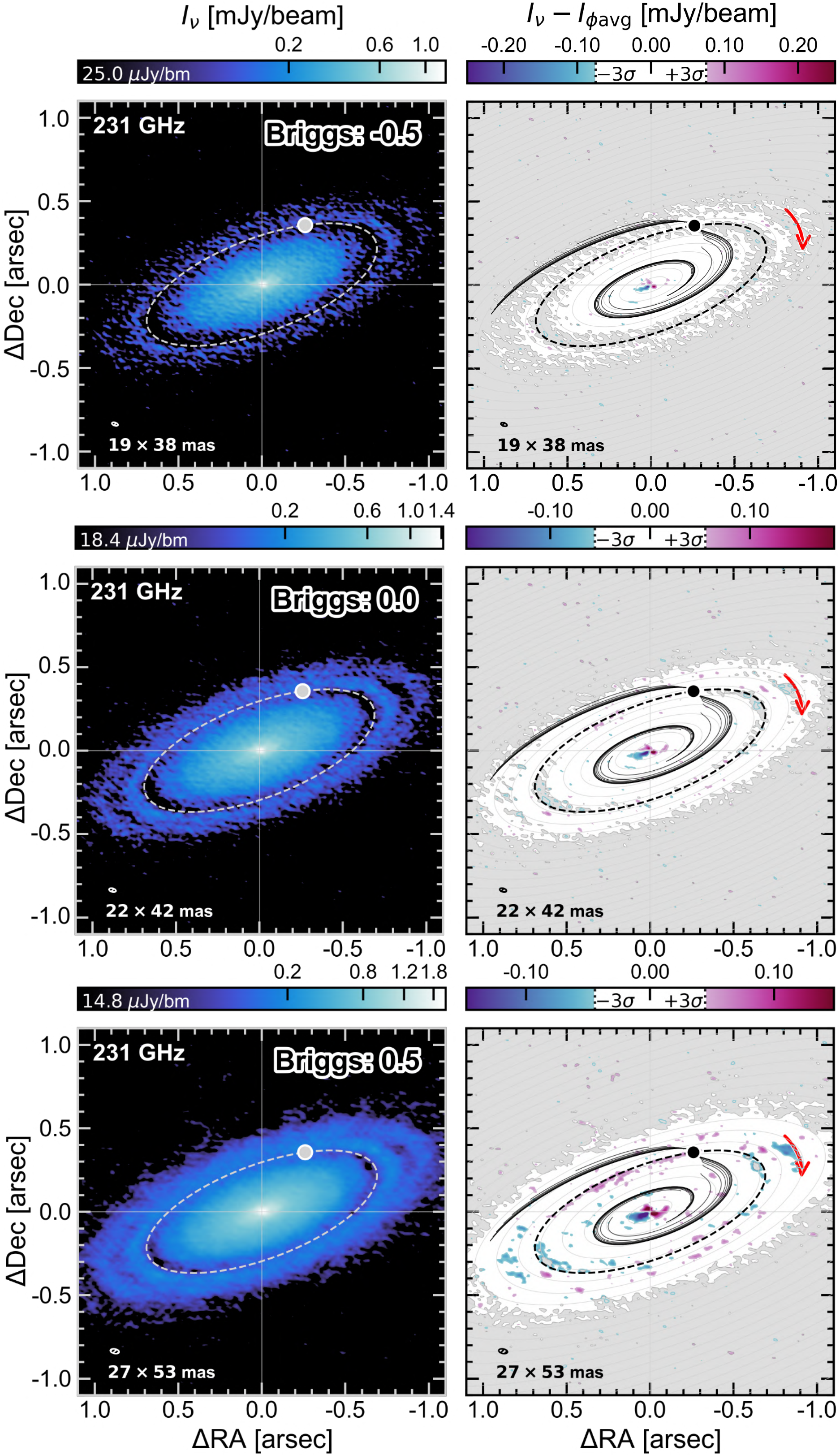}
\end{center}
\caption{\new{Subset of our re-imaging efforts with higher Briggs parameters to increase the beam width and sensitivity, showing the disk DoAr 25 as an example. The full set of images for DoAr 25, Sz 129, GW Lup, HD 163296 and HD 143006 for Briggs parameters $\in [-1, -0.5, -0.3, 0, 0.3, 0.5, 1, 2]$ is available at: \href{https://doi.org/10.6084/m9.figshare.21330426}{DOI: 10.6084/m9.figshare.21330426}.}
\label{fig:app:reimaging-DoAr25}}
\end{figure}

\new{\citet{speedie22-alma-dustspirals} argued that a beam size $\sim$2 times larger than the width of the spiral can yield a higher signal-to-noise detection in a residual map than a higher angular resolution image. Motivated by this, we re-imaged the publicly available DSHARP calibrated measurement sets of the 5 disks whose fiducial images yielded non-detections (DoAr 25, GW Lup, Sz 129, HD 163296 and HD 143006), varying the Briggs parameter to explore the full available range of angular resolutions and achievable sensitivities. We show an example of the results in Fig. \ref{fig:app:reimaging-DoAr25}. The full set is available at \href{https://doi.org/10.6084/m9.figshare.21330426}{DOI: 10.6084/m9.figshare.21330426}.}

\section{\new{Methods for Detecting Dust Spirals}} \label{app:frank-and-unsharp-masking}
\subsection{\new{Comparison with \texttt{frank}}} \label{app:subsec:frank-andrews}

\begin{figure}
\begin{center}
\includegraphics[width=8cm]{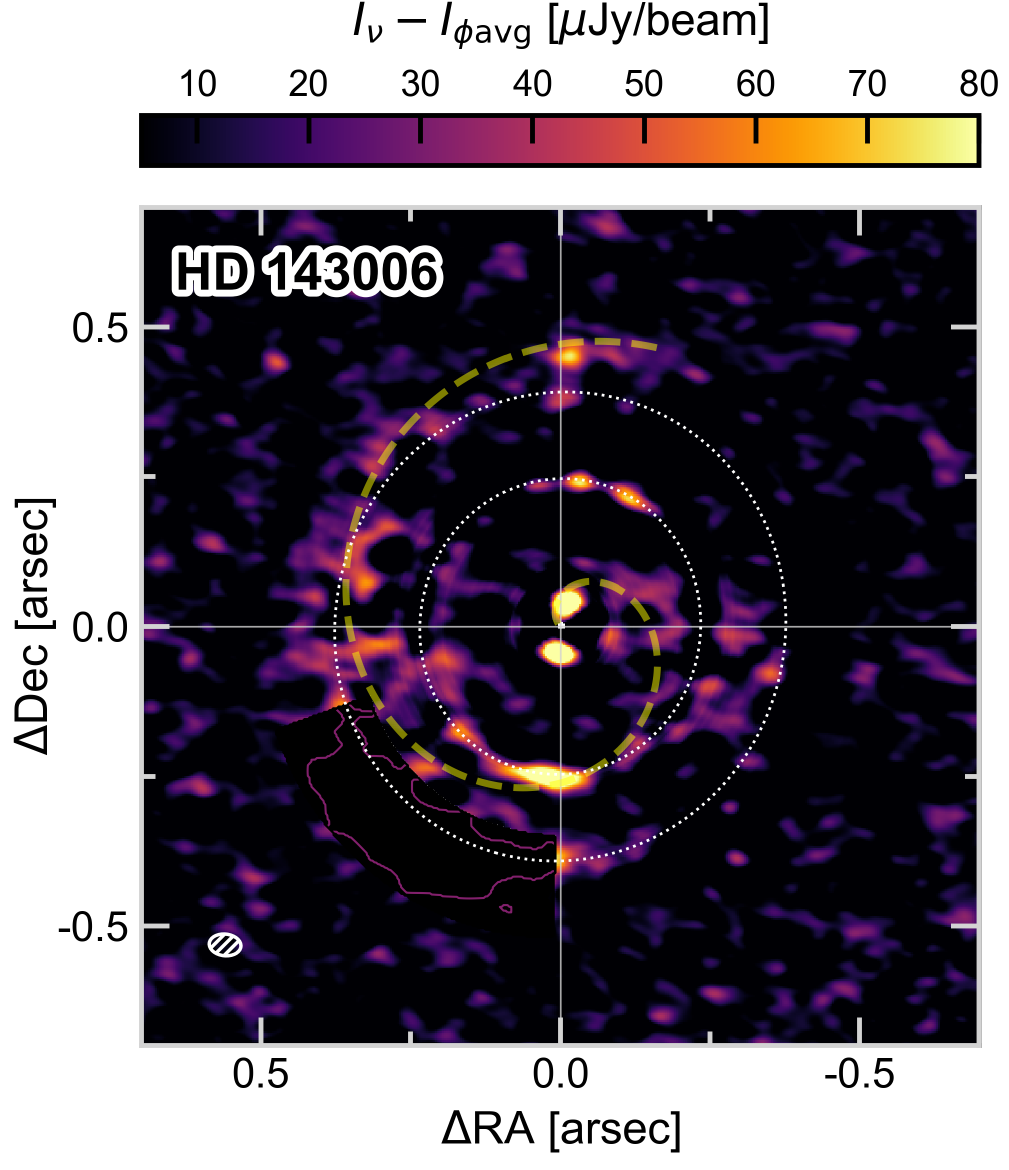}
\end{center}
\caption{\new{A re-presentation of our azimuthal average continuum residual map for HD 143006, for comparison to Fig. 4 of \citet{andrews21-frank-dsharp}. The dashed yellow curve is their (visually tuned, not fit) Archimedian spiral overlay, described in disk-frame coordinates by $R_{\rm spiral} = 0.170 + 0.067 \phi$ [arcsec]. The two dotted white ellipses mark the DSHARP-identified rings, B41 and B65.}
\label{fig:app:HD143006-andrews}}
\end{figure}

\new{All 8 of the DSHARP disks in our sample were analyzed \nuwe{in $uv$-space} with \texttt{frank} by \citet{jennings22-frank-dsharp}, and 4 of them also by \citet{andrews21-frank-dsharp}. 
Those works did not report detections of dust spirals in the \texttt{frank} residual maps of DoAr 25, GW Lup, Sz 129 and HD 163296, a result echoed by this letter. \nuwe{Side-by-side comparisons between the imaged \texttt{frank} residuals and our image-plane residuals shows the techniques give nearly identical results.}

\citet{andrews21-frank-dsharp} report a ``low-level'' tentative large-scale Archimedian spiral in HD 143006 (their Fig. 4). In our Fig. \ref{fig:app:HD143006-andrews}, we re-present the continuum residuals we obtained for HD 143006 (the exact same map as appears in the main text in the bottom right panel of Fig. \ref{fig:sec3:non-detections-2}), on the same colour scale as \citet{andrews21-frank-dsharp}, to more clearly demonstrate the extent to which our azimuthal average technique recovers this tentative spiral. The basic result is similar, 
and our method recovers the spotty residual features along the \citet{andrews21-frank-dsharp} spiral. \nuwe{In this case, some differences in the residuals can be attributed to the differing treatments of the confined azimuthal asymmetries.}
We note that the direction of this spiral (assuming it is trailing) 
implies counter-clockwise rotation for the HD 143006 disk, opposite to the clockwise rotation suggested by \citet{perezL18-hd143006}.} 

\subsection{\new{Additional Searches with the Unsharp Masking Method}} \label{app:high-pass}

\begin{figure*}
\begin{center}
\includegraphics[width=18cm]{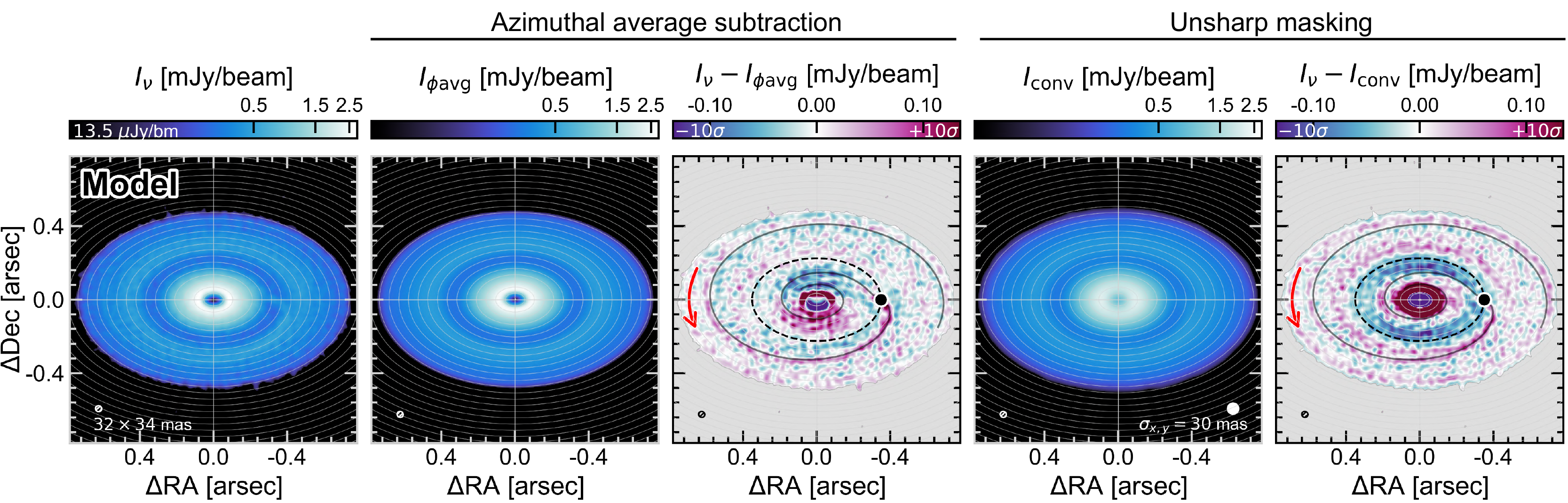}
\end{center}
\caption{\textbf{Comparing methods for detecting planet-driven dust spirals:} Azimuthal average subtraction vs. unsharp masking.
\textit{1st panel:} Synthetic continuum observation, $I_{\nu}$. 
\textit{2nd panel:} Background disk model $I_{\phi \, {\rm avg}}$ obtained by azimuthally averaging $I_{\nu}$, using knowledge of the disk inclination, position angle and phase offset. 
\textit{3rd panel:} Residual map as the difference between $I_{\nu}$ and $I_{\phi \, {\rm avg}}$. 
\textit{4th panel:} Background disk model $I_{\rm conv}$ obtained by convolving $I_{\nu}$ with a 2D Gaussian kernel of HWFM $\sigma_{x,y}$, making no assumptions on the disk geometry. Filled white circle represents the Gaussian kernel.
\textit{5th panel:} Residual map as the difference between $I_{\nu}$ and $I_{\rm conv}$. \\ 
In the residual maps, we overlay the theoretical prediction for the midplane spiral wake driven by the model planet \cite[][our Eqn. \ref{eqn:bae+zhu-spiral}]{bae+zhu18-spirals1,bae+zhu18-spirals2}, whose mass is $\Mp = 1.0 \, \Mth$. Only the dominant azimuthal mode ($m_{\rm dom}=({1\over2})(H/r)_{\rm p}^{-1}$) is shown, and the planet's outer spiral wake becomes more open than the predicted trajectory with distance from the planet.
\label{fig:app:high-pass-model}}
\end{figure*}

One of the most significant challenges to using the azimuthal average as a background model is that it makes assumptions on the disk geometry -- namely, that the dust disk is inherently axisymmetric and planar, and that one has accurate knowledge of the disk inclination and position angle. 
Artificial residuals can be introduced if one uses the incorrect disk geometry.
Creating residual maps with \texttt{frank} \citep{jennings20-frank} involves the same challenges \cite[e.g., Appendix A of both][]{jennings22-frank-taurus, andrews21-frank-dsharp}. 

In this section we explore an alternative technique that makes no assumptions on the disk geometry, and apply it to the disks that exhibited small- or large-scale azimuthal average residuals: HD 163296, HD 143006, and HD 97048. The technique has been referred to as ``unsharp masking'' \citep[e.g.,][]{perez16-elias27, meru17-elias27}, and involves convolving the observation with a normalized 2D Gaussian of HWFM $\sigma_{x,y}$ and subtracting the result from the original image. It is mathematically equivalent to the ``high-pass filtering'' technique  \citep[e.g.,][]{rosotti20-hd100453, norfolk22-hd100546} which involves suppressing large-scale spatial (angular) frequencies by convolution with a 1D Gaussian of HWFM $\sigma_{\nu}$ in the Fourier domain. We confirmed both give identical residual maps with the appropriately-scaled Gaussian kernels ($\sigma_{\nu} = 2\pi / \sigma_{x,y}$), but only show the former here so as to consistently work in the image plane.

Fig. \ref{fig:app:high-pass-model} compares the efficacy of the residual-making method we use in the main body of the paper to that of the unsharp masking method, using a synthetic continuum observation model\footnote{Downloadable from Figshare: \href{https://doi.org/10.6084/m9.figshare.19148912}{10.6084/m9.figshare.19148912}. This model contains a $1.0 \, Mth$ mass planet at 50 au in an adiabatic, slowly cooling ($\beta=10$), marginally optically thin ($\tau_0=0.3$) disk at a distance of 140 pc, observed with the C5+C8 ALMA configuration pair for an on-source time of 3.56 hours.} from \citet{speedie22-alma-dustspirals}. To produce the unsharp masking residual map, we convolve the model image ($I_{\nu}$, 1st panel) with a 2D Gaussian of HWFM $\sigma_{x,y}= 30$ mas ($I_{\rm conv}$, 4th panel) using \texttt{scipy.ndimage.gaussian\_filter}. \new{The residuals resulting from different kernel sizes will vary in morphology, as different kernels sequentially highlight different spatial scales of the image structures, and in practise we strongly encourage the observer to view the results for a range of $\sigma_{x,y}$. Here, we show the results for a select $\sigma_{x,y}$, chosen} such that $I_{\rm conv}$ is smoothed of blobs but still contains the radial structure present in $I_{\nu}$. This figure shows that the unsharp masking method can be effective, though it is prone to accentuating gaps and rings. 

\begin{figure*}
\begin{center}
\includegraphics[width=18cm]{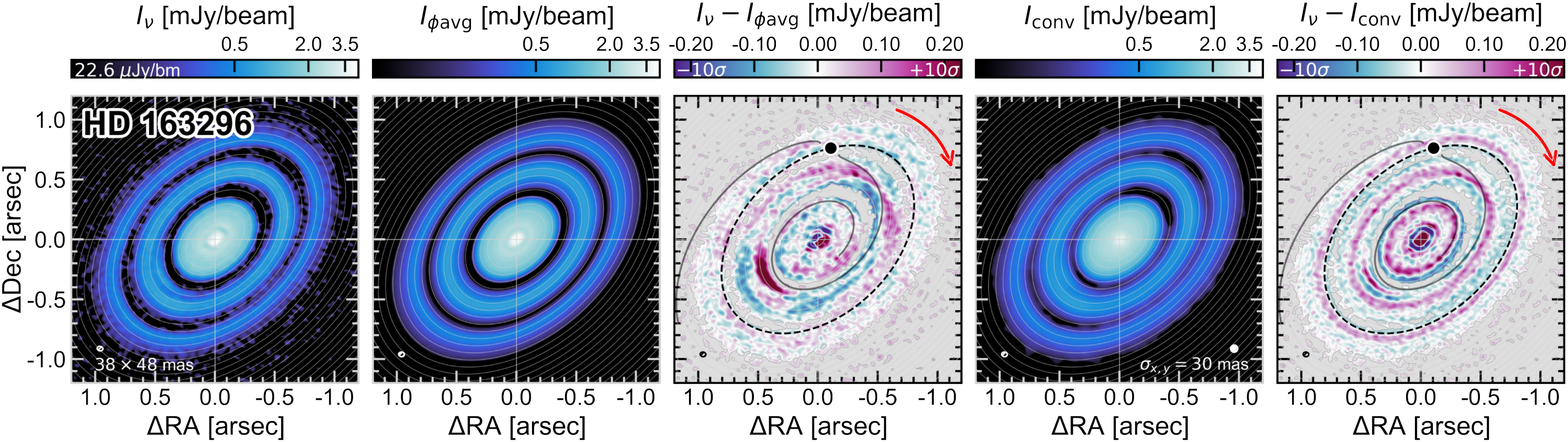}
\includegraphics[width=18cm]{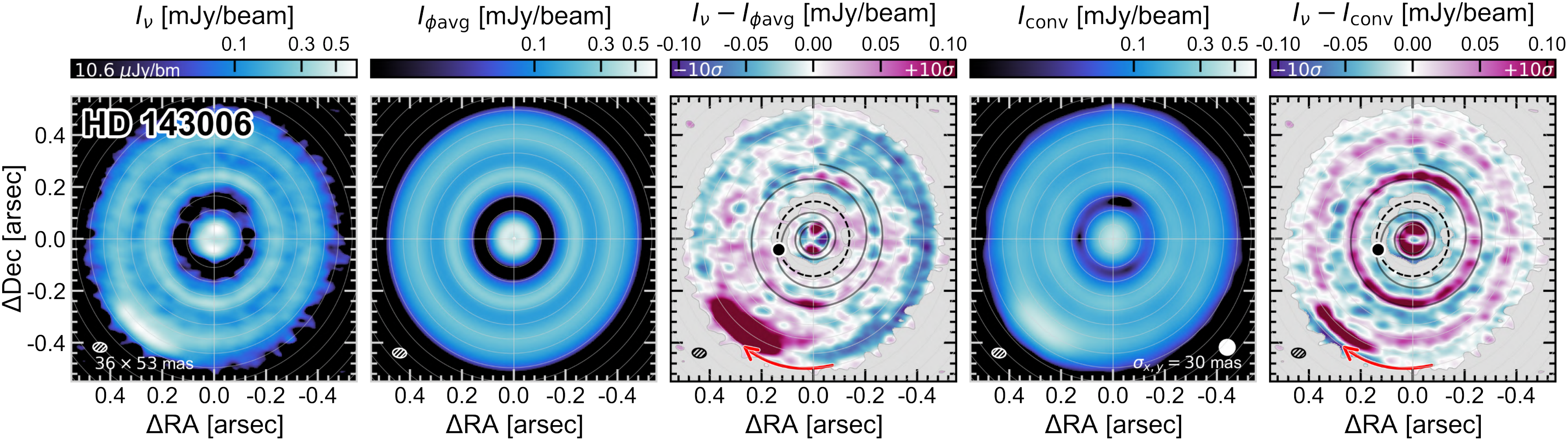}
\end{center}
\caption{\textbf{Additional searches for the predicted dust spirals:} HD 163296 and HD 143006. 
Panel layout similar to Fig. \ref{fig:app:high-pass-model}.
\label{fig:app:high-pass}}
\end{figure*}

\begin{figure*}
\begin{center}
\includegraphics[width=13cm]{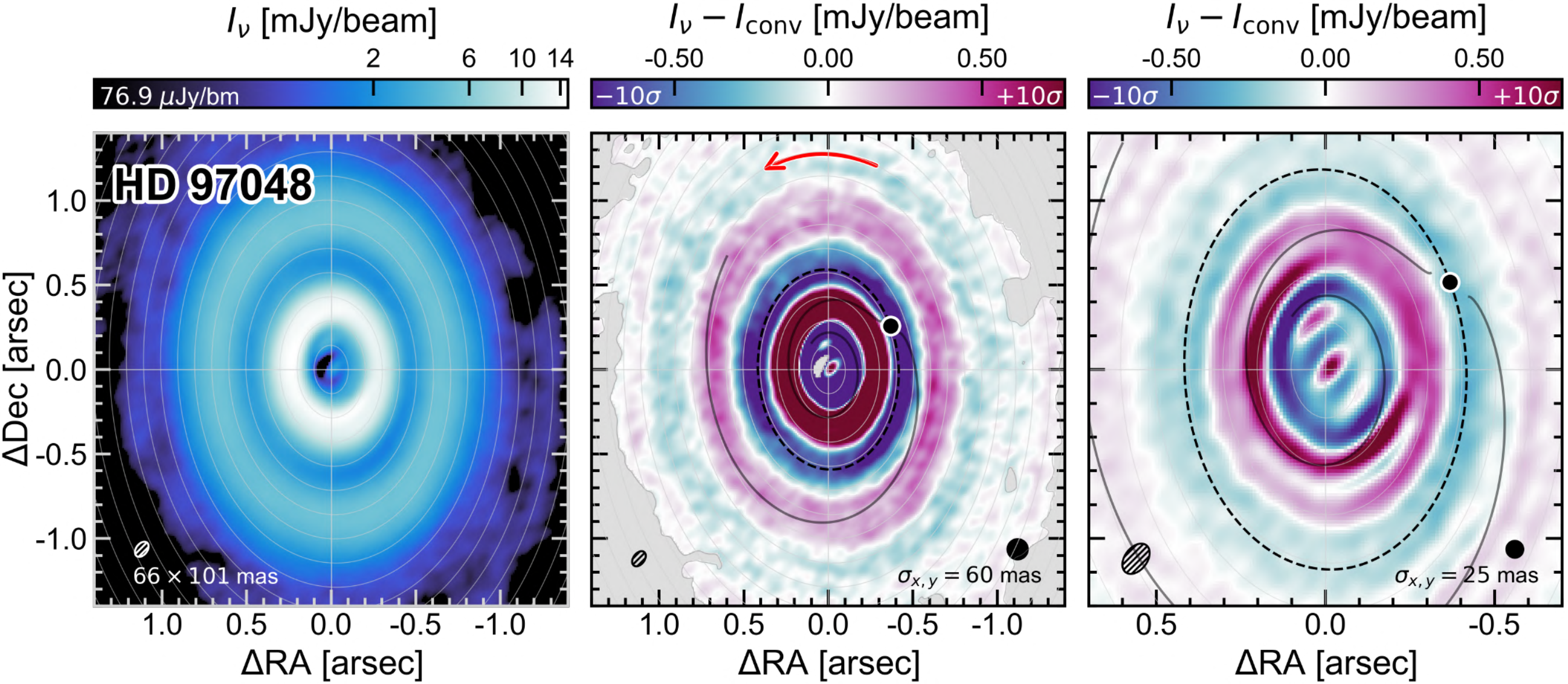}
\end{center}
\caption{\textbf{Additional searches for the predicted dust spirals:} HD 97048. Unsharp masking residuals produced with two 2D Gaussian kernels of different sizes to highlight substructure in the outer ring (\textit{2nd panel}) and inner ring (\textit{3rd panel}).
\label{fig:app:high-pass-hd97048}}
\end{figure*}

Nonetheless, for its main benefit of not requiring assumptions on the disk geometry, we apply it to the continuum observations of HD 163296, HD 143006 and HD 97048 in Figs. \ref{fig:app:high-pass} \& \ref{fig:app:high-pass-hd97048}. In the resulting residual maps for HD 163296 and HD 143006 (5th column of Fig. \ref{fig:app:high-pass}), we detect only ring and gap structures and no spirals, consistent with the azimuthal average method in the main text. In HD 97048 (Fig. \ref{fig:app:high-pass-hd97048}), we find that a single 2D Gaussian kernel does not highlight substructure in both the inner and outer ring simultaneously, and so we show the results for both a $\sigma_{x,y}= 60$ mas kernel and a $\sigma_{x,y}= 30$ mas kernel. The residual map produced with the larger kernel (2nd panel, Fig. \ref{fig:app:high-pass-hd97048}) shows no spiral structure in the outer ring. The residual map produced with the smaller kernel (3rd panel, Fig. \ref{fig:app:high-pass-hd97048}) reveals a double-ring structure over $\sim270^{\circ}$ of the inner ring, but no residuals that consistently follow the predicted trajectory for the planet's inner spiral arm.

\bibliography{bib}{}
\bibliographystyle{aasjournal}

\end{document}